\documentclass[%
 aip,
 amsmath,amssymb,
 reprint,%
]{revtex4-1}

\usepackage{graphicx}
\usepackage{dcolumn}
\usepackage{bm}

\usepackage[utf8]{inputenc}
\usepackage[T1]{fontenc}
\usepackage{mathptmx}
\usepackage{etoolbox}
\usepackage{float}
\usepackage{MnSymbol}
\usepackage{ulem}

\newcommand{\Sec}[1]{Sec.\,\ref{#1}}
\newcommand{\Eq}[1]{Eq.\,(\ref{#1})}
\newcommand{\Fig}[1]{Fig.\,\ref{#1}}

\newcommand{\RNum}[1]{\uppercase\expandafter{\romannumeral #1\relax}}
\usepackage{tikz}

\newcommand{\ie}{i.e.\,}
\newcommand{\au}{\,a.u.\,}

\makeatletter
\def\@email#1#2{%
 \endgroup
 \patchcmd{\titleblock@produce}
  {\frontmatter@RRAPformat}
  {\frontmatter@RRAPformat{\produce@RRAP{*#1\href{mailto:#2}{#2}}}\frontmatter@RRAPformat}
  {}{}
}%
\makeatother
\begin{document}

\preprint{AIP/123-QED}

\title{Stochastic resonance in vibrational polariton chemistry}
\author{Yaling Ke}
\email{yaling.ke@phys.chem.ethz.ch}
\affiliation{ 
Department of Chemistry and Applied Biosciences, ETH Zürich, 8093 Zürich, Switzerland
}%

\begin{abstract}
In this work, we systematically investigate the impact of ambient noise intensity on the rate modifications of ground-state chemical reactions in an optical cavity under vibrational strong-coupling conditions. To achieve this, we utilize a numerically exact open quantum system approach--the hierarchical equations of motion in twin space, combined with a flexible tree tensor network state solver. Our findings reveal a stochastic resonance phenomenon in cavity-modified chemical reactivities: an optimal reaction rate enhancement occurs at an intermediate noise level. In other words, this enhancement diminishes if ambient noise, sensed by the cavity-molecule system through cavity leakage, is either too weak or excessively strong. In the collective coupling regime, when the cavity is weakly damped, rate enhancement strengthens as more molecules couple to the cavity. In contrast, under strong cavity damping, reaction rates decline as the number of molecules grows.
\end{abstract}

\maketitle

\section{Introduction}
Placing reactive molecules within an optical microcavity under conditions of vibrational strong coupling (VSC) has recently shown promise for manipulating chemical reactions, such as cavity catalysis and cavity-induced reaction selectivities, even in the absence of light.\cite{Ebbesen_2016_ACR_p2403} This concept has been demonstrated in several experiments, which suggests that adjusting the distance between two reflective dielectric mirrors in a Fabry-P\'erot cavity so that the cavity frequency resonates with specific molecular vibrations can alter reaction rates or product ratios.\cite{Thomas_2016_ACE_p11634,Vergauwe_2019_ACIE_p15324,Lather_2019_ACIE_p10635,Thomas_2019_S_p615,Hiura_2019__p,Thomas_2020_PS_p249,Hirai_2020_C_p1981,Hirai_2020_ACE_p5370,Sau_2021_ACIE_p5712,Lather_2022_CS_p195,Ahn_2023_S_p1165,Ebbesen_2023__p} However, two independent experimental attempts to replicate these findings, while successfully reproducing VSC conditions, did not observe noticeable rate changes in an on-resonant cavity.\cite{imperatore2021reproducibility,wiesehan2021negligible} Different outcomes in the experiments suggest that additional factors beyond VSC, factors that may have been previously overlooked, could play a critical role in achieving cavity-induced modifications of ground-state chemical reactivities. In this study, we aim to explore one such experimental variable: the damping strength of the cavity mode.  This damping, caused by ambient noise surrounding the cavity due to the unavoidable cavity leakage, may influence the extent of cavity-induced rate modifications. 

Naturally, this intriguing yet controversial phenomenon has spurred a substantial amount of theoretical research.\cite{Galego_2016_NC_p13841,Galego_2017_PRL_p136001,Galego_2019_PRX_p21057,CamposGonzalezAngulo_2020_JCP_p,Mandal_2020_JPCL_p9215,Yang_2021_JPCL_p9531,Li_2021_JPCL_p6974,Li_2020_JCP_p234107,Li_2021_NC_p1315,Sun_2022_JPCL_p4441,Schaefer_2022_NC_p7817,Mandal_2022_JCP_p,Lindoy_2022_JPCL_p6580,Wang_2022_JPCL_p3317,Fischer_2022_JCP_p154305,Fiechter_2023_JPCL_p8261,Sokolovskii_2023_apa_p,Pavosevic_2023_NC_p2766} Recent studies using quantum dynamical simulations underscore the necessity of a quantum discrete state description for both molecular vibrations and the cavity mode.\cite{Lindoy_2023_NC_p2733,Lindoy_2024_N_p2617,Ying_2023_JCP_p84104,Hu_2023_JPCL_p11208,Ke_J.Chem.Phys._2024_p224704,Ke_2024_JCP_p54104} However, this requirement introduces significant challenges, particularly in the collective regime, where a large ensemble of molecules couples to the cavity mode. To address the computational challenges, we employ the hierarchical equations (HEOM) of motion along with a tree tensor network state (TTNS) solver.\cite{Ke_2023_JCP_p211102} The HEOM method is a well-established, numerically exact approach for open quantum system dynamics,\cite{Tanimura_1989_JPSJ_p101,Yan_2004_CPL_p216,Ishizaki_J.Phys.Soc.Jpn._2005_p3131,Xu_2007_PRE_p31107,Shi_2009_JCP_p84105,Yan_J.Chem.Phys._2014_p54105,Jin_J.Chem.Phys._2008_p234703,Schinabeck_Phys.Rev.B_2018_p235429,Hsieh_J.Chem.Phys._2018_p14103, Shi_J.Chem.Phys._2018_p174102,Tanimura_2020_JCP_p20901} enabling a non-perturbative and non-Markovian treatment of system dynamics and bath-related observables. The TTNS,\cite{Shi_Phys.Rev.A_2006_p22320,Tagliacozzo_Phys.Rev.B_2009_p235127,Murg_Phys.Rev.B_2010_p205105,Li_Phys.Rev.B_2012_p195137,Changlani_Phys.Rev.B_2013_p85107,Nakatani_J.Chem.Phys._2013_p134113,Lubich_SIAMJ.MatrixAnal.Appl._2013_p470,Murg_J.Chem.TheoryComput._2015_p1027,Gunst_J.Chem.TheoryComput._2018_p2026,Schroeder_Nat.Commun._2019_p1,Larsson_J.Chem.Phys._2019_p204102,Ferrari_Phys.Rev.B_2022_p214201,Montangero_Philos.Trans.R.Soc.A_2022_p20210065,Sulz_Phys.Rev.A_2024_p22420}  in addition, provides an efficient data compression scheme for storing and propagating the high-dimensional composite system-plus-bath wavefunction. 

To gain a preliminary understanding of how chemical reactions inside a cavity respond to ambient noise--stemming from interactions between the confined cavity mode and the continuum of far-field electromagnetic modes, which dampens the cavity's oscillatory dynamics--we begin our study on the single-molecule level. Our results show that reaction rate enhancements within the cavity exhibit a typical stochastic resonance feature\cite{Gammaitoni_1998_RMP_p223}: neither excessively weak nor overly strong damping of cavity dynamics supports effective cavity-induced rate modifications. Instead, the rate constant inside a resonant cavity reaches its peak at an intermediate level of cavity damping. In the collective coupling regime, where multiple molecules are interconnected through a shared coupling with the cavity mode, rate modifications can display distinct behaviors as the system size grows. Under weak ambient noise, the rate enhancement strengthens as more molecules are coupled to the cavity mode. On the contrary, under conditions of strong cavity damping, the rate enhancement is attenuated with the increased number of coupled molecules. These observations lead us to ask an essential question: Could an ensemble of molecules on the macroscopic scale, immersed in a solvent and thus exposed to significant background noise, act cooperatively through collective coupling to the cavity mode, and synergetically harness a feeble fluctuation to optimize the reaction efficiency?

The following sections present a detailed account of our findings. \Sec{sec:theory} outlines an open quantum system model that describes chemical reactions in a condensed-phase optical cavity and the quantum dynamics methodology used. \Sec{sec:results} presents and discusses the numerical results, followed by a summary and perspectives for our future research in \Sec{sec:conclusion}.

\section{\label{sec:theory}Theory}
We consider an open quantum system model to study chemical reactions inside an optical cavity, 
\begin{equation}
    H = H_{\rm S}+ H_{\rm E},
\end{equation}
where $H_{\rm S}$ describes the cavity-molecule system and $H_{\rm E}$ represents its surrounding environment. 

The reaction dynamics of $N_{\rm mol}$ molecules coupled to a single-mode cavity is described by the Pauli-Fierz light-matter Hamiltonian in the dipole gauge under the long-wavelength approximation \cite{Flick_2017_PotNAoS_p3026,Rokaj_2018_JPBAMOP_p34005,Mandal_2023_CR_p9786,Lindoy_2023_NC_p2733} 
\begin{equation}
\label{H_mol}
\begin{split}
    H_{\rm S} = &\overbrace{\frac{p_{\rm c}^2}{2} + \frac{1}{2} \omega_{\rm c}^2\left(x_{\rm c} + \sqrt{\frac{2}{\omega_{\rm c}}} \eta_{\rm c} \sum_{i=1}^{N_{\mathrm{mol}}} \vec{\mu}_i(x_i)\cdot \vec{e} \right)^2}^{H_{\rm c}(\mathbf{x})} 
    \\
    &+ \sum_{i=1}^{N_{\rm mol}}  \underbrace{\left[\frac{p_{i}^2}{2} + \frac{E_{\mathrm{b}}}{a^4}(x_i-a)^2(x_i+a)^2\right]}_{H_{\rm mol}^i}.
\end{split}
\end{equation}
The cavity mode is modeled as a harmonic oscillator with momentum $p_{\rm c}$, coordinate $q_{\rm c}$, and frequency $\omega_{\rm c}$. For convenience, we set $\hbar=1$ throughout this work. The light-matter coupling strength is characterized by the parameter $\eta_{\rm c}=\frac{1}{\omega_{\rm c}}\sqrt{\frac{\omega_{\rm c}}{2\epsilon_0V}}$, where $\epsilon_0$ is the permittivity of the medium within the cavity and $V$ is the quantization volume of the electromagnetic mode. The unit vector $\vec{e}$ points to the light polarization direction. Each molecule is represented by a reactive vibrational mode with mass-scaled momentum $p_{i}$, coordinate $x_{i}$, and a symmetric double-well potential energy surface with a barrier height $E_{\rm b}$ between two local minima at $x_i=\pm a$. The molecular dipole moment, denoted by $\vec{\mu}_i(x_i)$, depends on the reaction coordinate $x_i$.

\begin{figure*}
    \centering
        \begin{minipage}[c]{0.23\textwidth}
       \raggedright a) 
        \includegraphics[width=\textwidth]{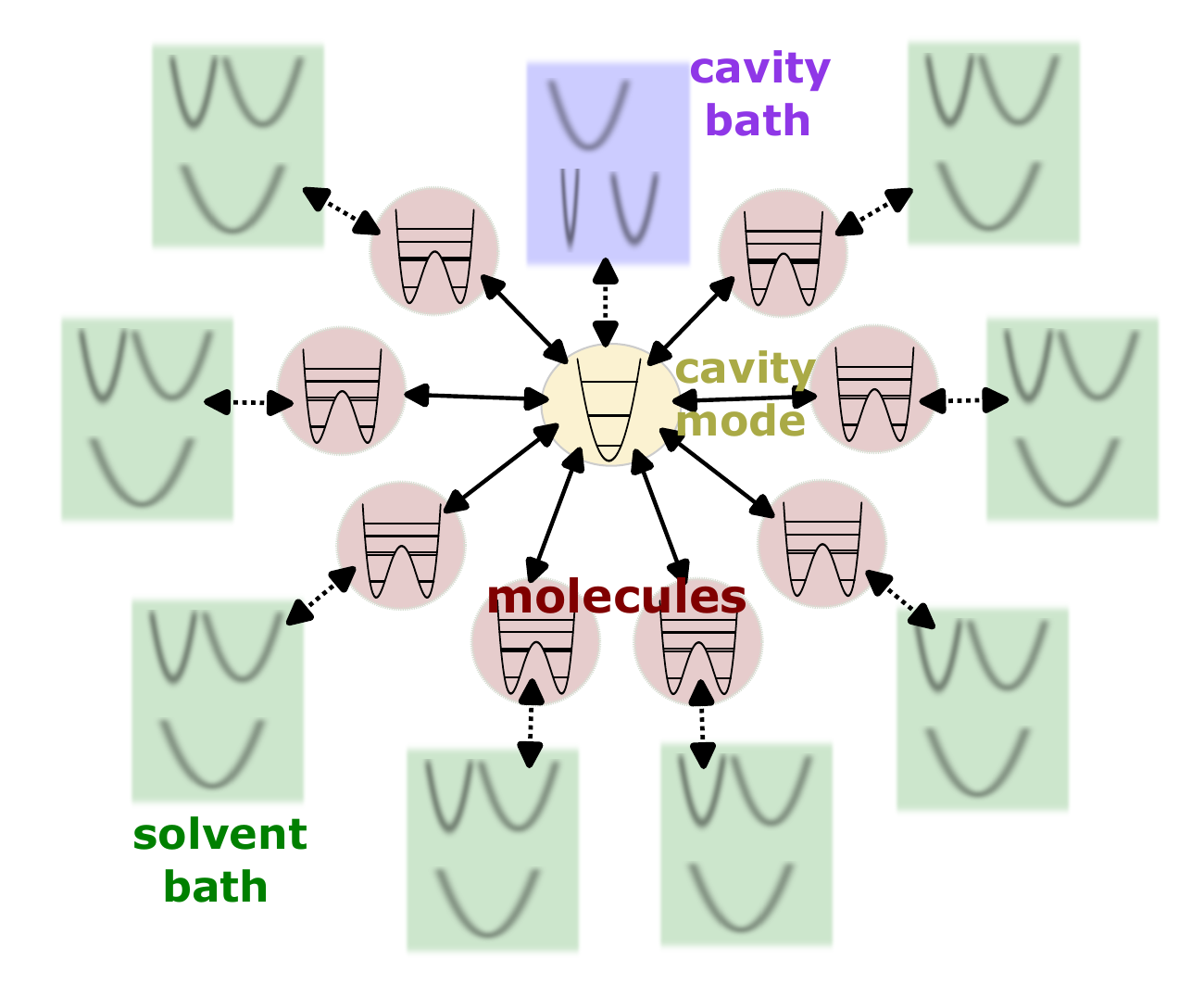}
    \end{minipage}
    \begin{minipage}[c]{0.25\textwidth}
       \raggedright b) 
        \includegraphics[width=\textwidth]{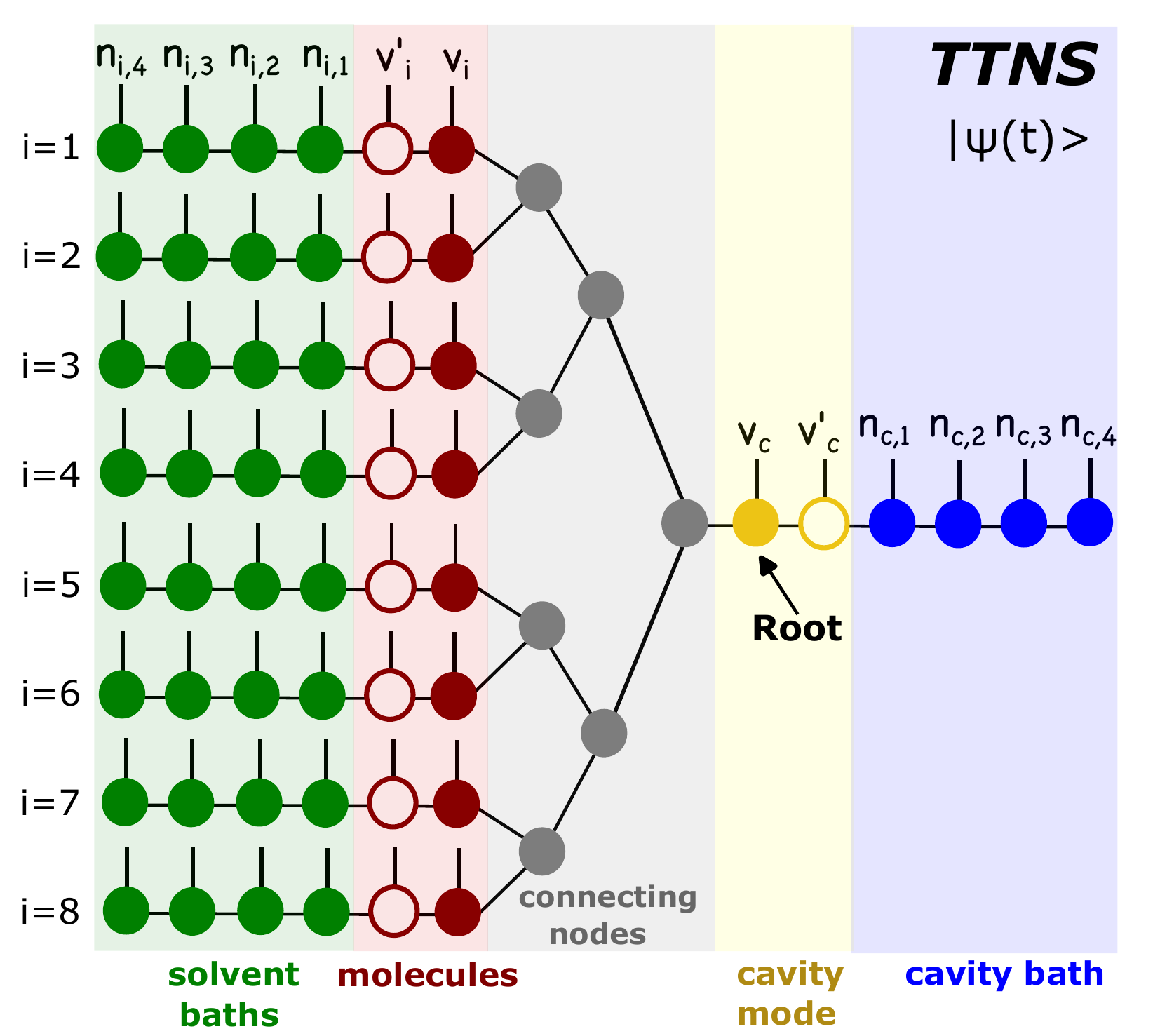}
    \end{minipage}
    \begin{minipage}[c]{0.25\textwidth}
       \raggedright c) 
        \includegraphics[width=\textwidth]{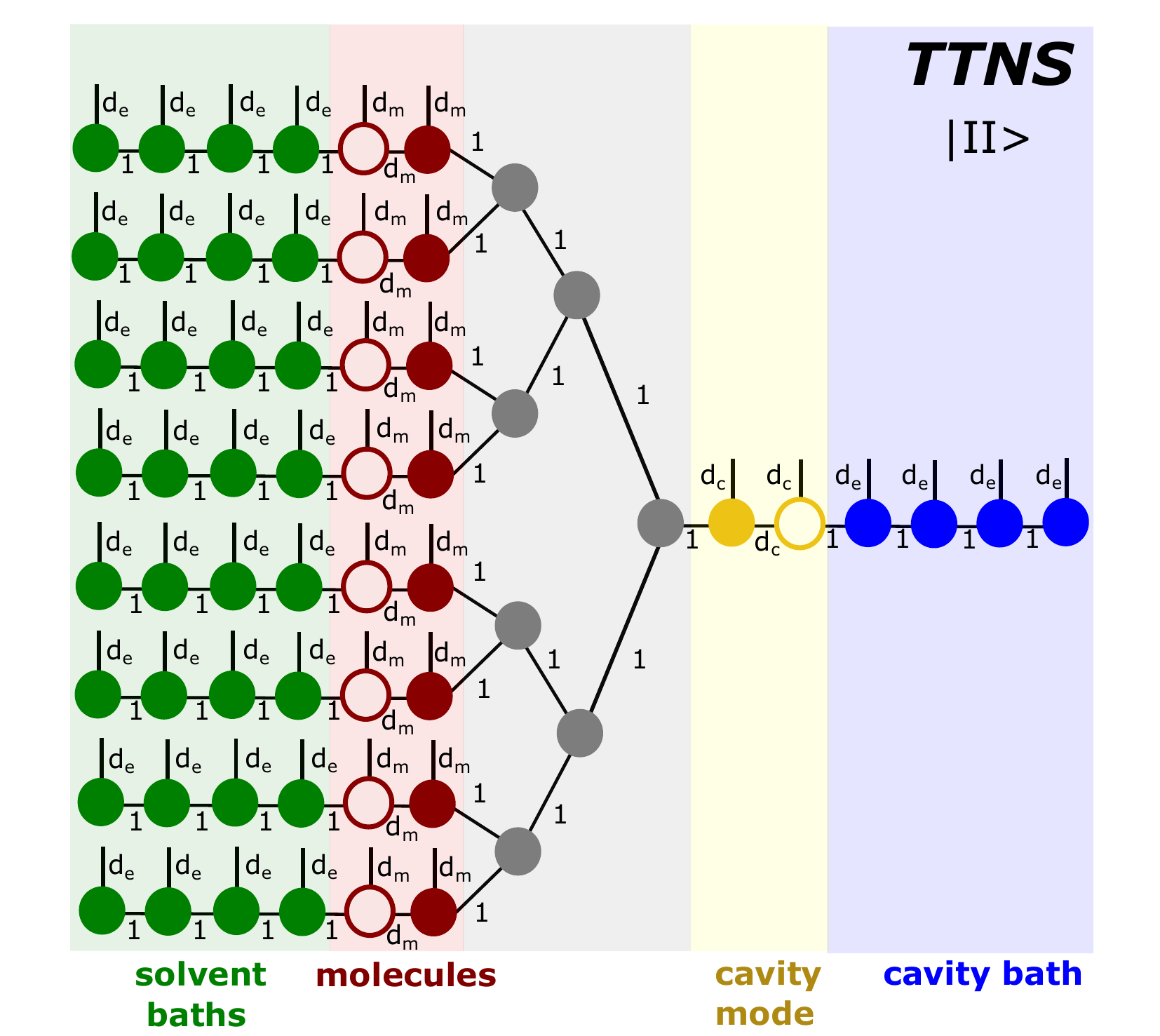}
    \end{minipage}    
    \begin{minipage}[c]{0.25\textwidth}
       \raggedright d)
        \includegraphics[width=\textwidth]{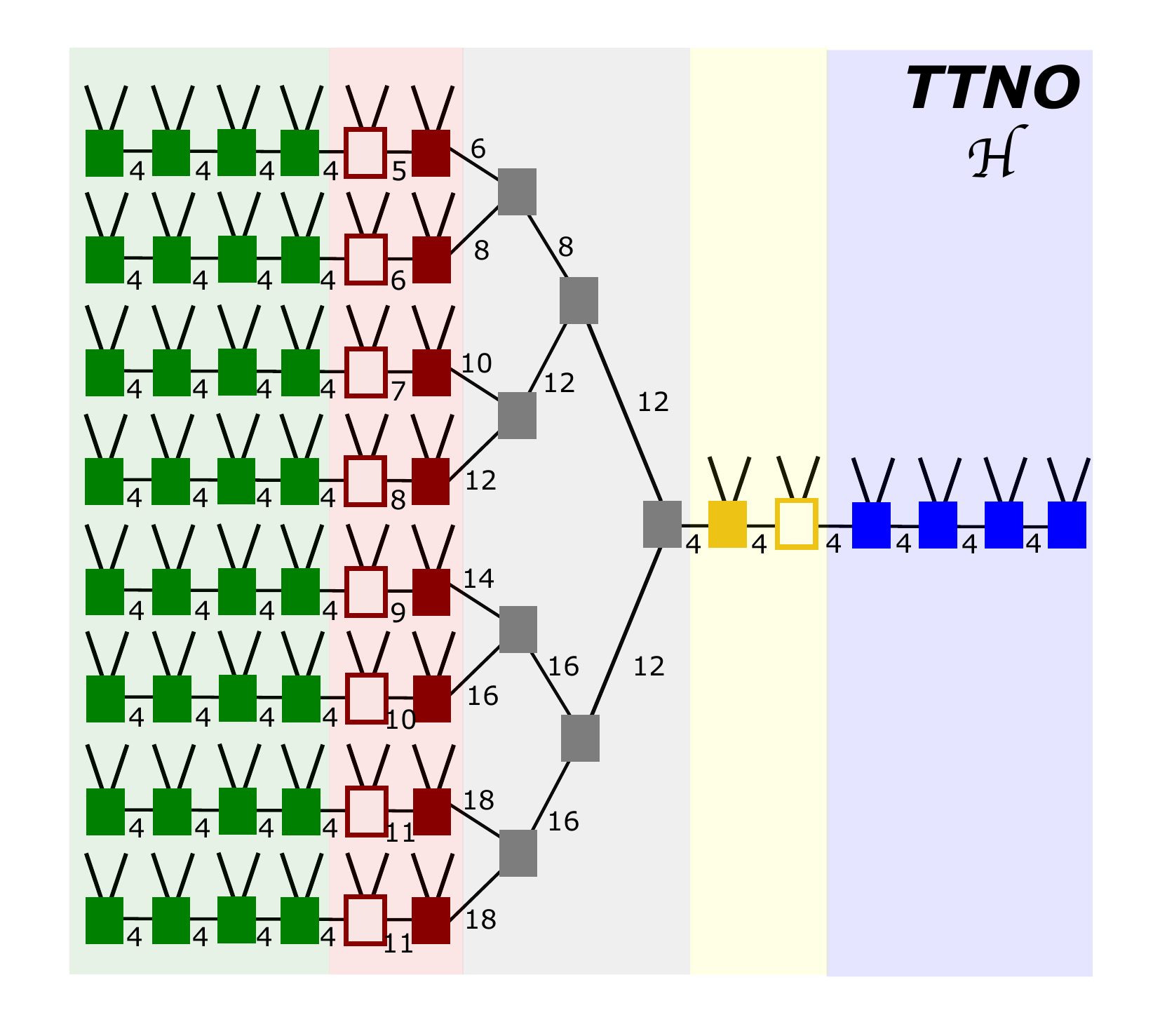}
    \end{minipage}
\caption{\label{model}
a) Schematic illustration of an open quantum system model for studying chemical reactions within an optical cavity. The system comprises a single-mode cavity field interacting with $N_{\rm mol}$ molecules. In the illustrated example, $N_{\rm mol}=8$. Each molecule is surrounded by a multitude of solvent molecules, while the confined cavity mode is also influenced by far-field radiation modes outside the cavity.   b) and c) are graphical representations of a TTNS decomposition of the extended wavefunction $|\Psi(t)\rangle$ and $|{\rm II}\rangle$, respectively. Each colored circular node in the TTNS represents a low-rank tensor associated with a specific component of the model: molecules (in red), solvent (in green), cavity mode (in yellow), and cavity bath (in blue). Each molecular node is linked to the cavity node through three connecting nodes shown in gray. d) represents a TTNO decomposition of the super-Hamiltonian $\mathcal{H}$. The square nodes represent the low-rank tensors in the TTNO, and they share the same color-coding scheme as in the TTNS. However, unlike in the TTNS representation, physical nodes in the TTNO have two dangling legs instead of one. The numbers labeled next to the connecting edges specify the bond dimensions of the virtual indices. The parameters $d_{\rm m}$, $d_{\rm c}$, and $d_{\rm e}$ indicate the number of basis functions for the vibrational, cavity, and effective environmental DoFs.}
\end{figure*}

The solvent and the continuum of electromagnetic modes together constitute the dissipative environment, with the Hamiltonian given by 
\begin{equation}
\label{H_sol}
\begin{split}
H_{\rm E} = & \sum_{k} \frac{P_{{\rm c}k}^2}{2}+ \frac{1}{2}\omega_{{\rm c}k}^2 \left(Q_{{\rm c}k}+\frac{g_{{\rm c}k}}{\omega_{{\rm c}k}^2}x_{\rm c}\right)^2  \\ 
&+ \sum_{i=1}^{N_{\rm mol}} \sum_k \frac{P^2_{ik}}{2}+\frac{1}{2}\omega_{ik}^2 \left(Q_{ik}+\frac{g_{ik}x_i}{\omega_{ik}^2}\right)^2,
\end{split}
\end{equation}
where the cavity mode and each molecule are coupled to their respective baths, represented as collections of harmonic oscillators, as schematically illustrated in \Fig{model}a. The bath associated with the cavity mode is referred to as the cavity bath, and each molecular bath is referred to as a solvent bath. The $k$th oscillator in bath $\alpha$ (either cavity or solvent) is described by its momentum $P_{\alpha k}$ and coordinate $Q_{\alpha k}$, with frequency $\omega_{\alpha k}$ and coupling strength $g_{\alpha k}$. The coupling to the system induces a displacement $g_{\alpha k}x_{\alpha}/\omega_{\alpha k}^2$ in each oscillator's coordinate. 

The environmental influence on the system dynamics is fully captured by the time-correlation function
\begin{equation}
\label{timecorrelation}
    C_{\alpha}(t) = \frac{1}{\pi} \int_{-\infty}^{\infty} \frac{e^{-i\omega t}}{1-e^{-\beta \omega}} J_{\alpha}(\omega) \mathrm{d}\omega=\sum_{p=1}^{P\rightarrow \infty} \lambda_{\alpha}^2\eta_{\alpha p} e^{-i\gamma_{\alpha p} t},
\end{equation}
where $\beta=1/k_{\rm B}T$ is the inverse temperature, and $J_{\alpha}(\omega)=\frac{\pi}{2}\sum_k \frac{c_{\alpha k}^2}{\omega_{\alpha k}}\delta(\omega -\omega_{\alpha k})$ is the spectral density function of bath  $\alpha$. This spectral density encodes the coupling-weighted density of states in the frequency domain for each bath. The time correlation function $C_{\alpha}(t)$ can be expanded analytically or numerically as a series of exponentials.\cite{Hu_2010_JCP_p101106,Xu_2022_PRL_p230601} 
This exponential expansion can be understood as mapping the original bath onto an effective one, which can be truncated efficiently at finite temperatures. In this effective representation, the bath consists of $P$ dissipative bosonic modes, each with a complex-valued frequency $\gamma_{\alpha p}$ and a component-wise coupling strength $\eta_{\alpha p}$ to the system. The reorganization energy $\lambda_{\alpha}^2=\sum_k \frac{g_{\alpha k}^2}{2\omega_{\alpha k}^2}$ quantifies the overall coupling strength between the system and bath $\alpha$. 

Rooted in the HEOM method in twin space for the system,\cite{Borrelli_2019_JCP_p234102,Borrelli_2021_WCMS_p1539,Ke_2022_JCP_p194102}  we can formulate a Schr\"odinger equation as
\begin{equation}
\label{Schroedinger}
    \frac{\mathrm{d}|\Psi(t)\rangle}{\mathrm{d} t} = -i\mathcal{H} |\Psi(t)\rangle
\end{equation}
for the extended wavefunction of the composite system
\begin{equation}
\label{Psi}
|\Psi(t)\rangle \equiv \sum_{\mathbf{n}} \sum_{v^{}_{\rm c} v'_{\rm c}\cdots v^{}_{N_{\rm mol}} v'_{N_{\rm mol}}  } C^{\mathbf{n}}_{v^{}_{\rm c}v'_{\rm c} \cdots v^{}_{N_{\rm mol}}v'_{N_{\rm mol}}}(t)|v^{}_{\rm c}v'_{\rm c}\cdots v^{}_{N_{\rm mol}}v'_{N_{\rm mol}}\rangle\otimes|\mathbf{n}\rangle.
\end{equation}
which includes both the system (cavity and molecules) and the effective baths. Each physical DoF in the system is represented by two independent indices, $v_{j}$ and $v'_{j}$, which arise from the purification of the density matrices. A single operator $O_{j}$ for the $j$th system DoF in Hilbert space corresponds to a pair of superoperators in twin space: $\hat{O}_{j}=O_{j}\otimes I_{j}$ and $\tilde{O}_{j}=I_{j}\otimes O_{j}^{\dagger}$, where $I_{j}$ is the identity operator for the $j$th DoF. Further, $|\mathbf{n}\rangle=|\cdots n_{\alpha p}\cdots\rangle$ represents the Fock state for the effective environmental modes in the number representation,  with creation and annihilation operators $b_{\alpha p}^{+}$ and $b_{\alpha p}$ defined as
\begin{subequations}
\begin{equation}
      b_{\alpha p}^{+} |\mathbf{n}\rangle = \sqrt{n_{\alpha p}+1}| \mathbf{n}_{\alpha p}^+ \rangle;
\end{equation}  
\begin{equation}
      b_{\alpha p} |\mathbf{n}\rangle = \sqrt{n_{\alpha p}}|\mathbf{n}_{\alpha p}^- \rangle,
\end{equation}
\end{subequations}
where $|\mathbf{n}_{\alpha p}^{\pm}\rangle=|\cdots n_{\alpha p}\pm 1 \cdots\rangle$. 
The super-Hamiltonian $\mathcal{H}$ is non-Hermitian and can be explicitly expressed as
\begin{equation}
\begin{split}
    \mathcal{H} = &\hat{H}_{\rm S}+\sum_{\alpha}\lambda_{\alpha}^2 \hat{x}_{\alpha}^2 -\tilde{H}_{\rm S}-\sum_{\alpha}\lambda_{\alpha}^2 \tilde{x}_{\alpha}^2 -i \sum_{\alpha}\sum_p  \gamma_{\alpha p} b_{\alpha p}^{+}b_{\alpha p}\\
    & +\sum_{\alpha}\sum_p \lambda_{\alpha}\left[(\hat{x}_{\alpha}-\tilde{x}_{\alpha})b_{\alpha p}     +(\eta_{\alpha p}\hat{x}_{\alpha}-\eta^*_{\alpha p}\tilde{x}_{\alpha})b_{\alpha p}^+ \right].
\end{split}
\end{equation}
By projecting \Eq{Schroedinger} onto a specific environmental Fock state, we obtain a differential equation for the auxiliary density operator, $\dot{\rho}^{\mathbf{n}}(t)=\langle \mathbf{n}|\dot{\Psi}(t)\rangle$, which reproduces the standard form of the HEOM method (see a review paper in Ref.\,\onlinecite{Tanimura_2020_JCP_p20901} and the literature therein). In particular, the reduced system density operator is given by $\rho_{\rm S}(t)=\rho^{\mathbf{0}}(t)=\langle \mathbf{n}=\mathbf{0}|\Psi(t)\rangle$. In this work, we assume that the system and the environment are initially factorized, so that $\rho(0) = \rho_{\rm S}(0) \cdot \rho_{\rm E}(0)$. Additionally, the environment is assumed to be in thermal equilibrium at the moment $t=0$. This assumption is equivalent to setting the environmental Fock state to the ground state at the initial time in the HEOM formalism.

To facilitate efficient propagation of the Schr\"odinger equation in \Eq{Schroedinger}, we employ a tree tensor network state solver.\cite{Ke_2023_JCP_p211102} The high-rank coefficient tensor $C^{\mathbf{n}}_{v^{}_{\rm c}v'_{\rm c}, \cdots v^{}_{N_{\rm mol}}v'_{N_{\rm mol}}}(t)$ is decomposed as a tree tensor network state, which is a powerful way to handle the complicated tensor structure involved in this system. As an example, a graphical representation of the TTNS decomposition of $|\Psi(t)$ for a system with $N_{\rm mol}=8$ molecules and $P=4$ effective environmental modes per bath is shown in \Fig{model}b. In this decomposition, a node with $z$-legs corresponds to a rank-$z$ tensor. To achieve an optimal balance between numerical stability, computational efficiency, and simulation time, the number of legs $z$ is constrained to $z\le 3$.\cite{Gunst_J.Chem.TheoryComput._2018_p2026} The physical indices $v^{}_{\rm c}/v'_{\rm c}$ (for the cavity), $v^{}_i/v'_i$ (for the molecules), and $n_{{\rm c}p},\,n_{ip}$ (for the effective environmental modes) in $|\Psi(t)\rangle$ are represented by the dangling legs in the yellow (filled/hollow), red (filled/hollow), blue, and green nodes, respectively. The connected legs between neighboring nodes $j$ and $k$ correspond to virtual indices $r_{jk}$, which runs from $1$ to $D_{jk}$. The maximum value among the bond dimensions $\{D_{jk}\}$ is called the maximal bond dimension $D_{\rm max}$. In practice, $D_{\rm max}$ is systematically increased until the results converge to a desired accuracy. The gray nodes, which have exclusively connected legs, are termed connecting nodes. The root node is assigned to the filled yellow node for the cavity, which has no parent node. All other nodes, except for the root node, have a parent node. This parent node is the adjacent node on the path pointing toward the root. In the binary tree structure displayed in \Fig{model}b, each node has at most two children. The leaves are the nodes without any offspring, representing the termination points of the tree.


The super-Hamiltonian $\mathcal{H}$ can be represented as a tree tensor network operator (TTNO) with the same tree topology as the TTNS used for $|\Psi(t)\rangle$. This decomposition ensures that when the TTNO is applied to the TTNS of $\Psi(t)$, the tree structure is preserved. The construction of the TTNO for $\mathcal{H}$ can be performed in an automatic manner.\cite{Milbradt_SciPostPhys.Core_2024_p36,Milbradt_arXivpreprintarXiv2407.13249_2024_p,Ceruti_arXivpreprintarXiv2405.09952_2024_p,Li_J.Chem.Phys._2024_p54116} In this work, we use an optimal TTNO constructor based on bipartite graph theory, which allows for an efficient construction of the TTNO.\cite{Li_J.Chem.Phys._2024_p54116} The resulting TTNO, corresponding to the tree shape shown in \Fig{model}b, is depicted in \Fig{model}d. A key difference between the TTNS and the TTNO is that in the TTNO, the physical nodes possess two dangling legs. To evolve \Eq{Schroedinger}, we adopt a time propagation scheme based on the time-dependent variational principle (TDVP),\cite{Haegeman_2016_PRB_p165116,Paeckel_2019_APN_p167998,Dunnett_2021_PRB_p214302,Bauernfeind_SciPostPhysics_2020_p24,Kloss_SciPostPhysics_2020_p70,Ceruti_SIAMJ.Numer.Anal._2021_p289} as detailed in Ref.\,\onlinecite{Ke_2023_JCP_p211102}. After solving \Eq{Schroedinger}, observables of interest can be extracted from the wavefunction $|\Psi(t)\rangle$. 

For the symmetric double-well model considered in this work, we assume that reactant and product regions are divided by a surface at $x_i^{\ddagger}=0$. The population of the $i$th molecule in the product region is obtained as
\begin{equation}
\label{product}
P_i(t) = \mathrm{Tr}\left\{  h_i\rho(t)\right\}  
    = \langle {\rm II}|\hat{h}_i|\Psi(t)\rangle.
\end{equation}
Here $h_i=\theta(x_i-x_i^{\ddagger})$ is the Heaviside projection operator onto the product region, where $x_i>x_i^{\ddagger}$. The wavefunction $|{\rm II}\rangle=|1_{\mathrm{sys}}\rangle\otimes |\mathbf{n=0}\rangle$ corresponds to the state where the system is in a unit vector state in twin space, \ie, $|1_{\mathrm{sys}}\rangle=\otimes_i\sum_{v^{}_i=v'_i}|v^{}_iv'_i\rangle$, and the environmental Fock state is in the ground state.

To perform the inner product in \Eq{product}, the wavefunction $|{\rm II}\rangle$ is also decomposed as a TTNS, maintaining the same tree topology as $|\Psi(t)\rangle$, as displayed in \Fig{model}c. The number next to the legs denotes the bond dimension of the corresponding index.  The parameters $d_{\rm m}$, $d_{\rm c}$, and $d_{\rm e}$ represent the number of basis functions for the molecular reactive DoF, cavity mode, and the effective environmental bosons, respectively. Each node in the TTNS is represented by a rank-$3$ tensor $V_{i_1\times i_2 \times i_3}$, where $i_1$, $i_2$, and $i_3$ are the dimensions of the respective indices. For the connecting nodes, the tensor is $V_{1\times 1\times 1}=1$. For the molecular (in red) or cavity (in yellow) nodes, the non-zero entries in the tensor are $V_{1\times d_{\rm m/c}\times d_{\rm m/c}}[1,j,k]=\delta_{jk}$ (filled node) and $V_{d_{\rm m/c}\times 1 \times d_{\rm m/c}}[j,1,k]=\delta_{jk}$ (hollow node). For the environmental nodes (in green and blue), the non-zero entry is $V_{1\times 1\times d_{\rm e}}[1,1,1]=1$. 

The rigorous expression for the first-order reaction rate constant of the $i$th molecule, moving from the reactant (left well) to the product region (right well), in the flux correlation function formalism is given by\cite{Miller_1983_JCP_p4889,Craig_2007_JCP_p144503,Chen_2009_JCP_p134505,Ke_2022_JCP_p34103}
\begin{equation}
\label{fluxcorrelationfunction}
k_i=\lim_{t\rightarrow t_{p}}k_i(t) = \lim_{t\rightarrow t_{p}}\frac{C_i^{\mathrm{fs}}(t)}{1-2P_i(t)}.
\end{equation}
Here, $t_p$ is the time at which $k_i(t)$ reaches a plateau. The flux-side correlation function is defined as
\begin{equation}
    C_i^{\mathrm{fs}}(t) = \mathrm{Tr}\{\rho(t) F_i\}= \langle {\rm II}|\tilde{F}_i|\Psi(t)\rangle,
\end{equation}
where $F_i = i [H, h_i]$ is the flux operator.

In the context of quantum dynamical simulations, it is important to note that the plateau value of $k_i(t)$ in \Eq {fluxcorrelationfunction}, which corresponds to the reaction rate, is independent of the choice of the initial density matrix. For the calculations in this work, the initial molecular vibrational wavefunction is largely prepared in the reactant region, and the cavity mode is assumed to be in thermal equilibrium. The system density operator is then defined as
\begin{equation}
\label{initialdensitymatrix}
    \rho_{\rm S}(0) = \left(\prod_{i=1}^{N_{\rm mol}} \frac{e^{-\frac{\beta H^i_{\rm mol}}{2}}(1-h_i)e^{-\frac{\beta H^i_{\rm mol}}{2}}}{Z_i}\right)\cdot\frac{e^{-\beta H_{\rm c}(\mathbf{x}=0)}}{\mathrm{Tr}\left\{e^{-\beta H_{\rm c}(\mathbf{x}=0)}\right\}}
\end{equation}
with $Z_i=\mathrm{Tr}_i\left\{e^{-\frac{\beta H^i_{\rm mol}}{2}}(1-h_i)e^{-\frac{\beta H^i_{\rm mol}}{2}}\right\}$ is a partition function for the $i$th molecule. The singular value decomposition of the initial density matrix in \Eq{initialdensitymatrix} is employed to construct the initial tree tensor network state. 

In the simulations conducted for this study, the following parameters are used for the double-well model, $E_b=2250\,\text{cm}^{-1}$ and $a=44.4$\au,  which are consistent with previous studies in the field.\cite{Lindoy_2023_NC_p2733,Lindoy_2024_N_p2617,Ying_2023_JCP_p84104,Hu_2023_JPCL_p11208,Ke_2022_JCP_p194102,Ke_2024_JCP_p54104,Fiechter_2023_JPCL_p8261} The Debye-Lorentzian spectral density function is adopted to characterize the baths and their interaction with the system
\begin{equation}
J_{\alpha} (\omega) = \frac{2\lambda_{\alpha}^2 \omega \Omega_{\alpha}}{\omega^2+\Omega_{\alpha}^2},
\end{equation}
 where $\Omega_{\alpha}$ is the characteristic frequency of bath $\alpha$. All solvent baths share the same parameters, which are fixed as $\lambda_i=100\,\text{cm}^{-1}$ and $\Omega_i=200\,\text{cm}^{-1}$. The exponential expansion of the time correlation function in \Eq{timecorrelation} is performed using the Pad\'e decomposition scheme.\cite{Hu_2010_JCP_p101106} The molecular vibrational DoFs are described by the potential-optimized discrete variable representation,\cite{Colbert_1992_JCP_p1982,Echave_1992_CPL_p225} with the lowest $d_{\rm m}$ eigenstates taken into account. For simplicity, the primary analysis assumes that all molecular dipoles are oriented along the direction of the cavity field polarization, $\vec{\mu}_i(x_i)\cdot 
  \vec{e}=x_i$. The influence of other dipole orientations, including oppositely and anisotropically aligned configurations, is addressed comprehensively in the supplementary information (SI). The cavity mode is represented by the lowest $d_{\rm c}$ harmonic eigenstates. The convergence is meticulously verified, with careful attention to the following parameters: $d_{\rm m}$, $d_{\rm c}$, $d_{\rm e}$, time step $\Delta t$, the number of Pad\'e poles $P$, and the maximal bond dimension $D_{\rm max}$. Without further specification, we set $d_{\rm m}=6$, $d_{\rm e}=10$, $\Delta t=0.5\,\text{fs}$, $P=4$, and $D_{\rm max}=30$. Depending on the specific parameters, the timescale for $k(t)$ to reach a plateau, $t_{p}$, varies between $10\,\text{ps}$ and $50\,\text{ps}$.

It is worth noting that while the matrix product state (MPS),\cite{Schollwoeck_2011_APN_p96} also known as tensor train (TT),\cite{Oseledets_2011_SJSC_p2295} is a particular, simpler case of the TTNS, using MPS/TT for simulating reaction dynamics in vibrational polariton chemistry--especially in the collective regime--is computationally inefficient. This inefficiency arises due to the complexity of correlations in such systems, which require more flexible tensor network topologies. Further details on the numerical performance of different tensor network structures are provided in Appendix A. 

\begin{figure}
    \centering
        \begin{minipage}[c]{0.4\textwidth}
       \raggedright a) $\eta_c=0.00125$\au \\
        \includegraphics[width=\textwidth]{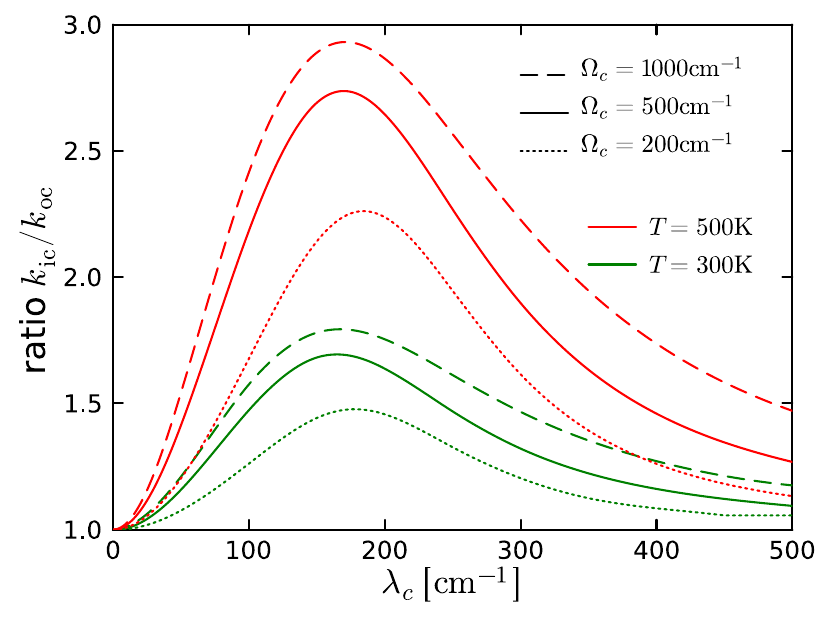}
    \end{minipage}
    \begin{minipage}[c]{0.4\textwidth}
       \raggedright b) $\eta_c=0.005$\au \\
        \includegraphics[width=\textwidth]{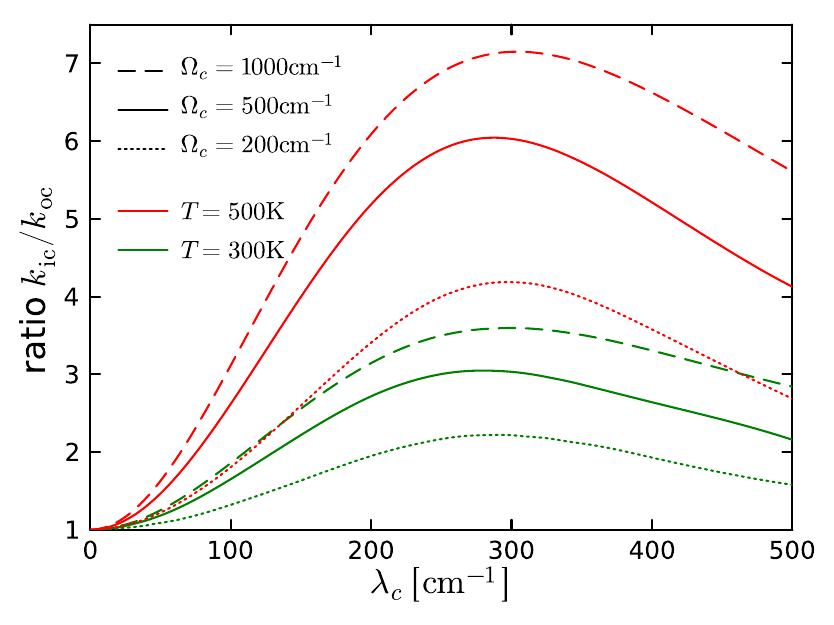}
    \end{minipage}
\caption{\label{monomer}
Ratio of rates $k_{\rm ic}/k_{\rm oc}$ for a single molecule undergoing a reaction inside an optical cavity $(k_{\rm ic})$ compared to the outside rate $(k_{\rm oc})$ as a function of the cavity damping strength $\lambda_{\rm c}$. The results for two different temperatures $T$ and three characteristic frequencies $\Omega_{\rm c}$ of the cavity bath, and light-matter coupling strengths ($\eta_{\rm c}=0.00125$\au in the upper panel and $\eta_{\rm c}=0.005$\au in the lower panel) are shown. The cavity frequency is kept at $\omega_{\rm c}=1185\,\text{cm}^{-1}$ and $d_{\rm c}=10$.  }
\end{figure}

\section{\label{sec:results}Results}
In practice, the cavity is not perfectly reflective and is therefore subject to external noise from the continuum of far-field electromagnetic modes. Previous numerically exact quantum dynamical studies in the single-molecule limit have demonstrated that, in a resonant but lossless cavity, reaction rates remain unaffected.\cite{Lindoy_2023_NC_p2733,Ying_2023_JCP_p84104,Ke_J.Chem.Phys._2024_p224704}  Furthermore, an interpolated rate theory at the single-molecule level, based on Fermi's Golden Rule in both the lossless and lossy limits, which is valid in the Markovian and weak light-matter coupling regime, predicts a turnover in reaction rates inside the cavity as a function of the cavity loss rate.\cite{Ying_2024_CM_p110} These findings give an important hint that cavity leakage may be a crucial factor in altering reaction dynamics inside an optical microcavity. Accordingly, the objective of this work is to understand how ground-state chemical reactions within an optical cavity are influenced by cavity damping. We begin by investigating the effect of cavity damping strength on reaction rate modifications at the single-molecule level, then extend our analysis to the collective regime, where multiple molecules are simultaneously coupled to the cavity mode.

In this work, cavity leakage is modeled by introducing a cavity bath, with the cavity damping strength quantified by the parameter $\lambda_{\rm c}$. This is verified by examining the dynamics of the average photonic excitation number, given by $\langle n_{\rm c}(t) \rangle=\mathrm{Tr}\{a^{\dagger}_{\rm c}a_{\rm c}\rho(t)\}$. Here, $a^{\dagger}_{\rm c}=\sqrt{\omega_{\rm c}/2}(x_{\rm c}-ip_{\rm c}/\omega_{\rm c})$ and $a_{\rm c}=\sqrt{\omega_{\rm c}/2}(x_{\rm c}+ip_{\rm c}/\omega_{\rm c})$ represent the creation and annihilation operators of the confined cavity mode, respectively. Detailed data supporting this verification are provided in the SI.

\begin{figure}
    \centering
        \begin{minipage}[c]{0.4\textwidth}
       \raggedright a) 
        \includegraphics[width=\textwidth]{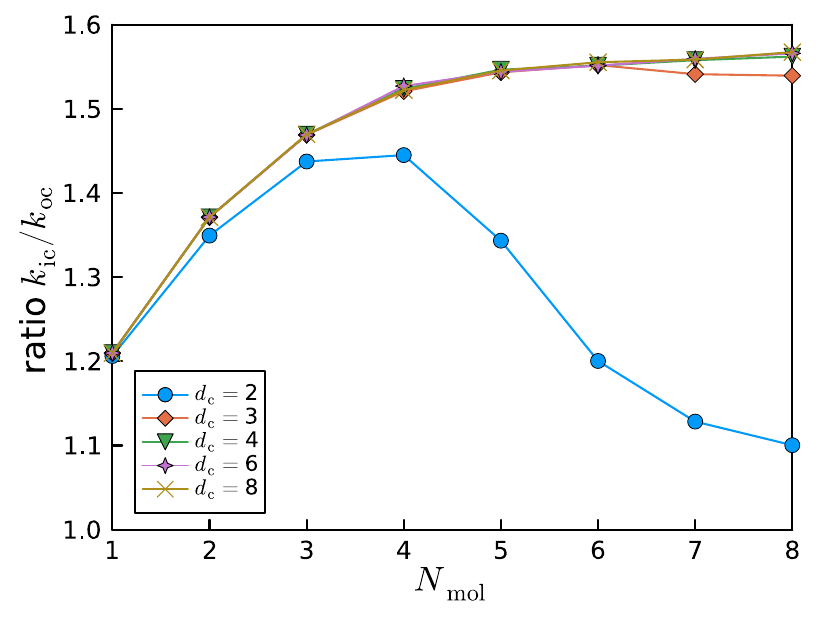}
    \end{minipage}
    \begin{minipage}[c]{0.4\textwidth}
       \raggedright b) 
        \includegraphics[width=\textwidth]{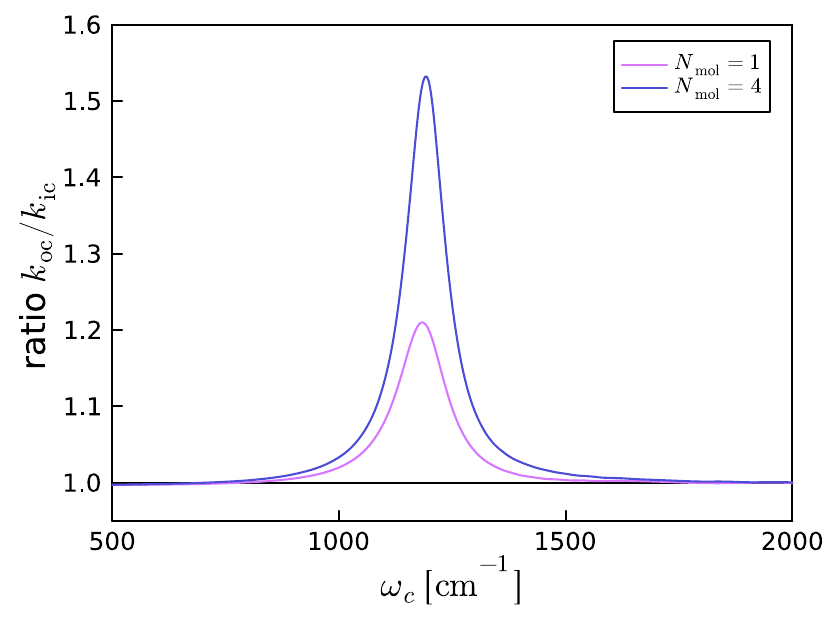}
    \end{minipage}
\caption{\label{weakdamping}
a) Ratio of rates $k_{\rm ic}/k_{\rm oc}$ as a function of the number of molecules collectively coupled to the cavity mode. Note that $k_{\rm ic}/k_{\rm oc}$ is identical for each individual molecule as the constituent of the molecular ensemble. Results are shown for different numbers of $d_{\rm c}$ lowest photonic states. The cavity frequency is fixed at $\omega_c=1185\,\text{cm}^{-1}$. b) Rate modification profile as a function of the cavity frequency for cases with one and four molecules coupled to the cavity mode, respectively, and with $d_{\rm c}=6$. Remaining parameters are $\lambda_{\rm c}=50\,\text{cm}^{-1}$, $\Omega_{\rm c}=1000\,\text{cm}^{-1}$, $\eta_{\rm c}=0.00125$\au, and $T=300\,\text{K}$.}
\end{figure}
\begin{figure}
    \centering
        \begin{minipage}[c]{0.4\textwidth}
       \raggedright a) 
        \includegraphics[width=\textwidth]{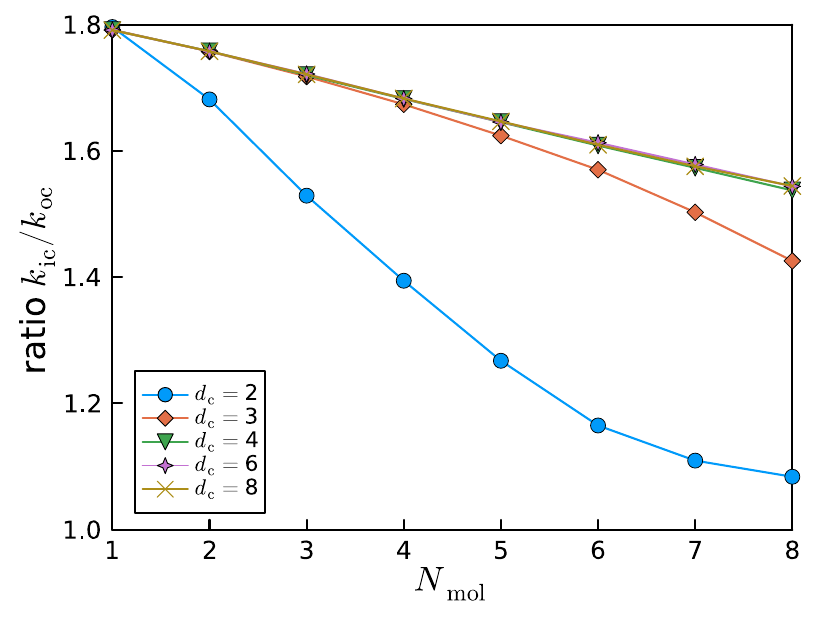}
    \end{minipage}
    \begin{minipage}[c]{0.4\textwidth}
       \raggedright b) 
        \includegraphics[width=\textwidth]{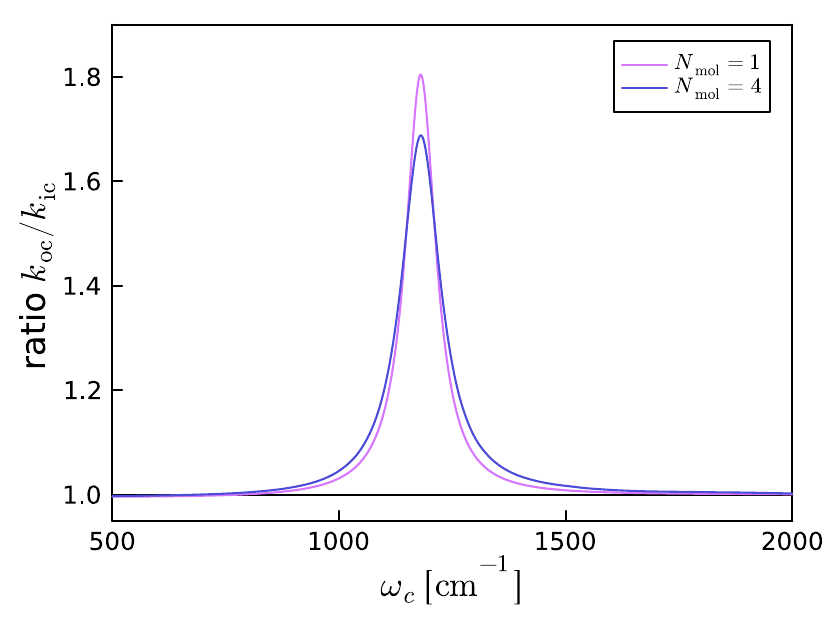}
    \end{minipage}
\caption{\label{strongdamping}
Same as \Fig{weakdamping} except that $\lambda_{\rm c}=165\,\text{cm}^{-1}$.}
\end{figure}

\Fig{monomer} presents the cavity-induced rate modification, represented as $k_{\rm ic}/k_{\rm oc}$--the ratio of reaction rates inside and outside the cavity--in the single-molecule limit, plotted as a function of $\lambda_{\rm c}$. The cavity frequency is set at $\omega_{\rm c}=1185\,\text{cm}^{-1}$, closely resonant with molecular vibrational transitions and corresponding to the peak in the rate modification profile over $\omega_{\rm c}$ (see also \Fig{weakdamping}b and \Fig{strongdamping}b). The rate within the cavity initially increases with $\lambda_{\rm c}$, reaching a maximum before this rate enhancement diminishes as $\lambda_{\rm c}$ continues to increase. This behavior is examined with various values of temperature $T$, the characteristic frequency $\Omega_{\rm c}$ of the cavity bath, and the light-matter coupling strength $\eta_{\rm c}$. The extent of the rate modification depends on temperature, $\Omega_{\rm c}$, and $\eta_{\rm c}$. Notably, the cavity-induced rate modification is a finite-temperature effect, vanishing at low temperatures and saturating as the temperature rises, as demonstrated in our previous study.\cite{Ke_J.Chem.Phys._2024_p224704} Furthermore, the rate modification initially increases with $\Omega_{\rm c}$, then plateaus, and gradually decreases at higher $\Omega_{\mathrm{c}}$ values, as demonstrated in the SI. However, both temperature $T$ and $\Omega_{\rm c}$ only slightly affect the crossover value of $\lambda_{\rm c}$. In contrast, increasing the light-matter coupling strength $\eta_{\rm c}$ not only strengthens the rate modification significantly but also shifts the crossover $\lambda_{\rm c}$ to a much higher value, rendering the rate modification more resilient to ambient noise. It is worth noting that, in the strong light-matter coupling regime, the inclusion of higher photonic excitation states is critical. Restricting the consideration to only the photonic ground state and first excited state leads not only to incorrect predictions of the resonant peak in the rate modification profile as a function of the cavity frequency $\omega_{\mathrm{c}}$, but also to a significant overestimation of the turnover $\lambda_{\mathrm{c}}$. The detailed analysis supporting these findings and a more systematic study of the impact of the light-matter coupling strength on reaction dynamics are provided in the SI.

This observation fits into the notion of the stochastic resonance theory, which was first proposed to explain the periodic recurrence of Earth's ice ages.\cite{Benzi_1981_JPAMG_p453,Nicolis_1981_SP_p473} Over time, it has prospered into a broad field on its own with diverse applications across physics, chemistry, biomedical sciences, engineering, and beyond, highlighting the constructive role of an optimal amount of noise.\cite{Wiesenfeld_1995_N_p33,Gammaitoni_1998_RMP_p223,Ando2000,Haenggi_2002_C_p285,Harmer_2002_ITIM_p299,Wellens_2003_RPP_p45,Moss_2004_CN_p267}  Stochastic resonance refers to a phenomenon where a moderate level of noise improves a system's sensitivity to detect weak, information-carrying signals.\cite{Gammaitoni_1998_RMP_p223} For chemical reactions inside an optical cavity, three key elements have been possessed for the emergence of stochastic resonance. Firstly, reactions take place in a solvent, which serves as a source of substantial background noise. Secondly, reactive molecules are inherently nonlinear systems that support threshold-like barrier-crossing events. Thirdly, cavity leakage exposes reactions to a generically weak coherent input, which could boost the system's sensitivity, depending on the noise level.

In many cases, stochastic resonance is observed as a collective effect.\cite{Jung_1992_PRA_p1709,Lindner_1995_PRL_p3} For example, it is found that the sensitivity of paddlefish electroreceptors to detect individual plankton is enhanced when plankton form a swarm.\cite{Russell_1999_N_p291} Similarly, it is of great interest to explore how the reaction dynamics inside the cavity change in the collective regime, by gradually increasing the number of molecules coupled to the cavity mode. 

We begin with a weakly-damped cavity mode characterized by $\lambda_{\rm c}=50\,\text{cm}^{-1}$, with other parameters set as $T=300\,\text{K}$ and $\Omega_{\rm c}=1000\,\text{cm}^{-1}$. In addition, the light-matter coupling strength per molecule is fixed at $\eta_{\rm c}=0.00125$\au, meaning that the Rabi splitting $\Omega_R$ increases as the number of molecules grows. For a homogeneous molecular aggregate in the all-aligned dipole configuration, the rate ratio $k_{\rm ic}/k_{\rm oc}$ is identical for every individual molecule in the ensemble. In \Fig{weakdamping}a, we show $k_{\rm ic}/k_{\rm oc}$ as a function of the number of molecules collectively coupled to the cavity mode, with $d_{\rm c}$ lowest photonic states included in the calculations. As more molecules are coupled through the cavity mode, the reaction rate of an individual molecule as part of the connected molecular network inside the cavity increases, in agreement with the finding in Ref.\,\onlinecite{Lindoy_2024_N_p2617}. It is worth noting that, in the collective regime, the involvement of higher photonic excitation manifolds also becomes necessary for accurate results. For the scenario in \Fig{weakdamping}, at least the two lowest photonic excited states $(d_{\rm c}=3)$ are required to reach convergence for $N_{\mathrm{mol}}\geq2$, and $d_{\rm c}=4$ for $N_{\mathrm{mol}}\geq 7$. If only the lowest photonic excited state $(d_{\rm c}=2)$ is considered, both the reaction rates and the decay of oscillations in the flux-side correlation functions and $\langle n_{\rm c}(t) \rangle$ are significantly underestimated, as demonstrated in the SI, suggesting the importance of multi-photon processes in capturing the dissipative nature of the cavity bath, which is crucial for cavity-induced reaction modifications. Furthermore, we observed that as the number of molecules increases from $N_{\rm mol}=1$ to $N_{\rm mol}=4$, the height of the resonant peak in the reaction rate modification profile increases, as shown in \Fig{weakdamping}b, while the peak becomes sharper. Specifically, the full width at half maximum (FWHM) is reduced from approximately $125\,\text{cm}^{-1}$ for $N_{\rm mol}=1$  to $105\,\text{cm}^{-1}$ for $N_{\rm mol}=4$.

In contrast, when the cavity mode is more heavily damped with $\lambda_{\rm c}=165\,\text{cm}^{-1}$ while all other parameters remain unchanged, increasing $N_{\rm mol}$ results in a steady drop of $k_{\rm ic}/k_{\rm oc}$, as shown in \Fig{strongdamping}a. Additionally, in this scenario, the FWHM of the resonant peak in the cavity-induced rate modification profile broadens from about $80\,\text{cm}^{-1}$ for $N_{\rm mol}=1$  to $105\,\text{cm}^{-1}$ for $N_{\rm mol}=4$, as shown in \Fig{strongdamping}b. Moreover, with the larger $\lambda_{\rm c}$, more photonic excited states contribute to the collective reaction dynamics within an optical cavity. For instance, $d_{\rm c}=4$ photonic states are required for $N_{\rm mol}>4$, as illustrated in \Fig{strongdamping}a.

These findings support the hypothesis that cavity-induced rate modifications might also be influenced by factors beyond just the presence of VSC, typically indicated by a large Rabi splitting. Specifically, in the single-molecule limit, the rate modification profile as a function of the cavity frequency $\omega_{\mathrm{c}}$ and the absorption spectra (both detailed in the SI) reveal that a sharp resonant peak in the rate modification profile can emerge regardless of whether the Rabi splitting is present or absent. In the VSC regime--characterized by a Rabi splitting that significantly exceeds the full width at half maximum of the split peaks in the absorption spectra--only a broad and weak resonant peak is observed.
Our results demonstrate the reaction rate is considerably impacted by the cavity damping strength, which is highly associated with the specifics of the cavity setup, including the material composition and quality factor. In the Markovian regime, it is known that the Rabi splitting $\Omega_R$ is proportional to the square root of molecular concentration $\sqrt{N_{\rm mol}/V}$.\cite{Ebbesen_2016_ACR_p2403} In the collective coupling scenario, where the Rabi splitting increases with the number of molecules coupled to the cavity mode as we maintain a constant light-matter coupling strength per molecule, the rate changes exhibit diverse trends depending on the cavity damping strength. Further results detailing the ratio of $k_{\mathrm{ic}}/k_{\mathrm{oc}}$ as a function of $N_{\mathrm{mol}}$ for various cavity damping strength $\lambda_{\mathrm{c}}$ are provided in the SI. These observations suggest that the Rabi splitting alone is not the determining factor for observing rate modification within the cavity. The possibility that the experimental reproducibility challenges in polaritonic chemistry might be related to the delicacy of stochastic resonance was first hinted in Refs.\,\onlinecite{sidler2022perspective, ruggenthaler2023understanding}. Our findings in this work may provide an important numerical clue into discrepancies in experimental outcomes.\cite{Hiura_2019__p,Lather_2019_ACIE_p10635,imperatore2021reproducibility,wiesehan2021negligible}  

We also notice that in the collective coupling regime, while the reaction rate varies with $N_{\rm mol}$, it does not exceed the maximum rate observed at the single-molecule level under the same conditions of temperature, $\Omega_{\rm c}$, and $\eta_{\rm c}$. This suggests that studies at the single-molecule level may provide valuable predictions on the upper limit of cavity-induced rate modification for a specific chemical reaction in the collective coupling regime. 

From another perspective, our results clearly show that the reaction rate for a single molecule, as the building block of a molecular network, can differ significantly from that of an isolated molecule. Moreover, the reaction kinetics of a molecular assembly organized by the cavity mode, in response to a certain level of external noise, are highly sensitive to the system's size.  This raises a compelling question: For a macroscopically large ensemble of molecules collectively coupled to a cavity mode, could even a very weak perturbation surrounding the cavity be perceived and amplified throughout the entire connected cavity-molecule system, leading to a drastic change in reaction dynamics? Validating this conjecture will require further investigations, both theoretically and experimentally.

\section{\label{sec:conclusion}Conclusion}
In summary, we conduct a quantum dynamical and numerically exact study on the impact of cavity damping strength on condensed-phase chemical reactions inside an optical microcavity under VSC conditions, using the HEOM method with an efficient tree tensor network decomposition scheme. Our results reveal that cavity-induced rate modifications exhibit a typical stochastic resonance feature. Specifically, reaction rates within a cavity are significantly enhanced when the cavity frequency is tuned in resonance with the molecular vibrational transition energy, not only under VSC conditions but also at an intermediate level of cavity damping. Additionally, distinct behaviors are observed as the number of molecules collectively coupled to the cavity mode increases, depending on the cavity damping strength: under weak damping, reaction rates inside the cavity can be boosted as the number of molecules grows, whereas strong damping leads to a decrease in reaction rates with molecular aggregation.

We hope this study will inspire deeper curiosity into the intricacies of vibrational polariton chemistry experiments, and into factors beyond VSC that could potentially lead to major ramifications in the reaction dynamics. It is important to note, however, that a substantial gap remains between the current model study and experimental observations. Looking forward, we plan to develop a more advanced tensor network state approach that extends beyond the tree tensor network topology, allowing us to investigate more realistic multi-mode cavities. We also aim to extend this approach to realistic molecular systems by combining it with the ab initio simulations, moving beyond simplified models to more accurately represent experimental conditions.

\begin{figure*}
    \centering
        \begin{minipage}[c]{\textwidth}
       \raggedright a) \\
        \includegraphics[width=0.96\textwidth]{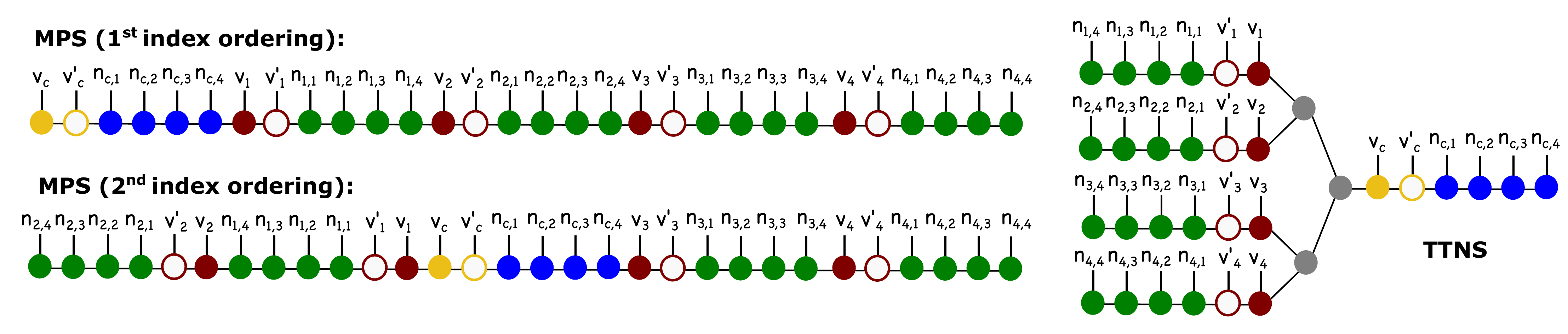}
    \end{minipage} 
    \begin{minipage}[c]{0.32\textwidth}
       \raggedright b) MPS ($1^{\rm st}$ index ordering) \\
        \includegraphics[width=\textwidth]{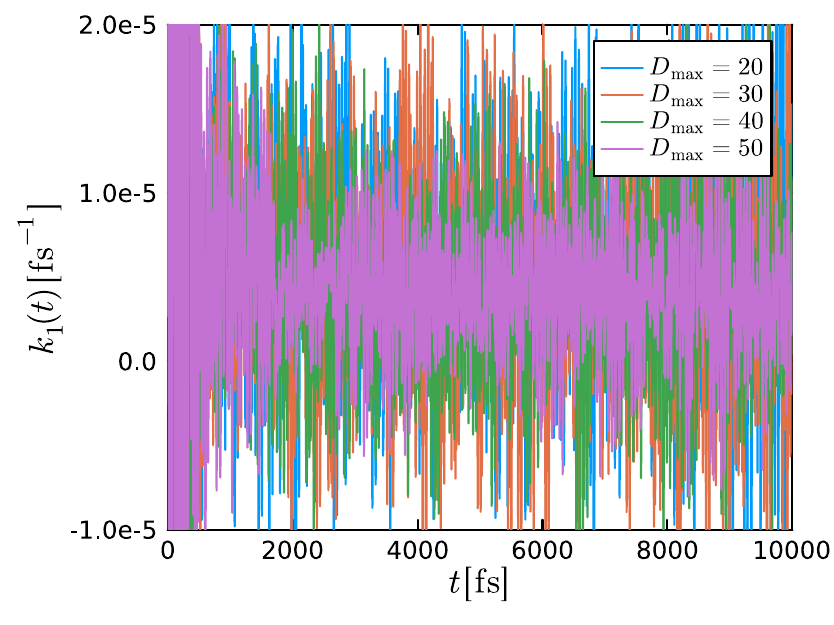}
    \end{minipage}
        \begin{minipage}[c]{0.32\textwidth}
       \raggedright c) MPS ($2^{\rm nd}$ index ordering) \\
        \includegraphics[width=\textwidth]{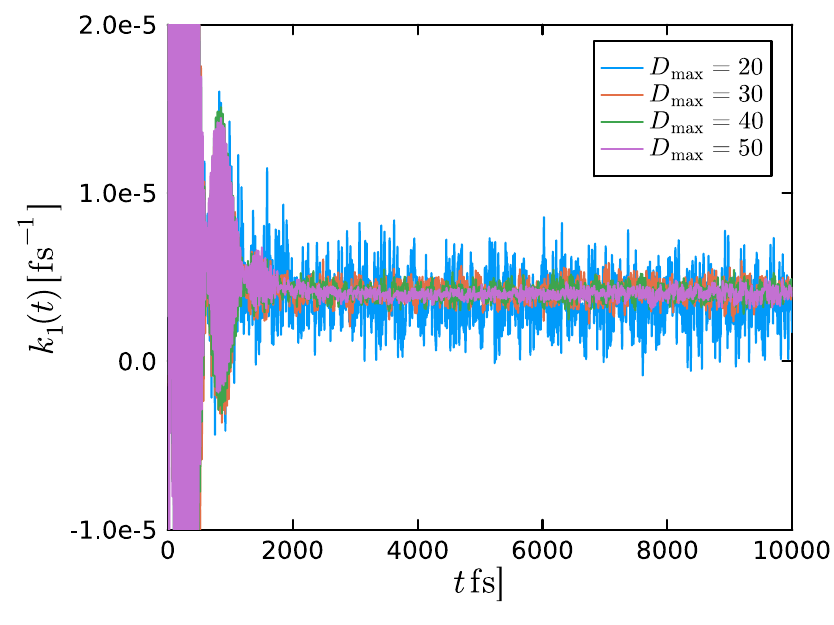}
    \end{minipage}
    \begin{minipage}[c]{0.32\textwidth}
       \raggedright d) TTNS
        \includegraphics[width=\textwidth]{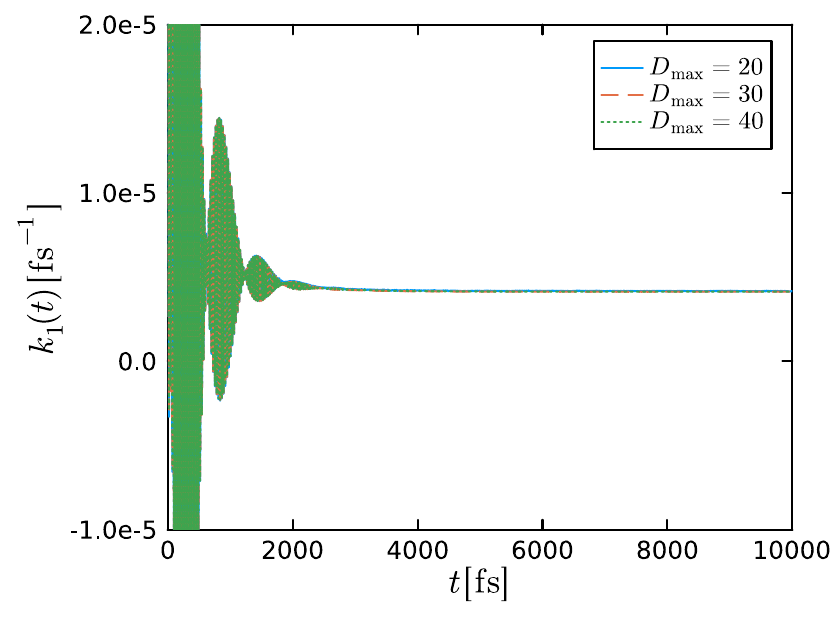}
    \end{minipage}
\caption{\label{MPSvsTTNS}
a) Illustration of MPS/TT decompositions with two different arrangements of physical indices on the left side, and a TTNS decomposition on the right side of the extended wavefunction $|\Psi(t)\rangle$ for $N_{\rm mol}=4$. b-d) Rescaled flux-side correlation function $k_1(t)$ for the first molecule, as defined in \Eq{fluxcorrelationfunction}, is shown with varying maximal bond dimensions. Each panel corresponds to a tensor network state decomposition scheme illustrated in a). Results in c) for $D_{\rm max}=20$, $D_{\rm max}=30$, and $D_{\rm max}=40$ are represented by solid, dashed, and dotted lines, respectively, to avoid visual overlap from superimposition.  The pertinent parameters are set as follows:  $\omega_{\rm c}=1185\,\text{cm}^{-1}$, $\lambda_{\rm c}=165\,\text{cm}^{-1}$, $\Omega_{\rm c}=1000\,\text{cm}^{-1}$, $\eta_{\rm c}=0.00125$\au, $d_{\rm c}=6$, and $T=300\,\text{K}$. }
\end{figure*}

\begin{acknowledgments}
The author appreciates valuable discussions with Richard Milbradt on the algorithm for constructing a TTNO with state diagrams, and with Prof. Jiushu Shao on the stochastic resonance theory. Y.K. thanks the Swiss National Science Foundation for
the award of a research fellowship. 
\end{acknowledgments}

\section*{Supplementary information}
See the supplementary material for the details and further analysis of (1) the impact of cavity bath parameters $\lambda_{\mathrm{c}}$ and $\Omega_{\mathrm{c}}$ on the damping of system dynamics; (2) the impact of bath characteristic frequency $\Omega_{\alpha}$ on reaction dynamics; (3) the impact of the light-matter coupling strength on reaction dynamics; (4) infrared absorption line shapes and rate modification profiles in the single-molecule limit; (5) the importance of higher photonic excitation manifold; (6) the impact of various noise level in collective coupling regime; and (7) the impact of dipole orientations. 

\section*{Data Availability Statement}
The data and code that support the findings of this work are available from the corresponding author upon reasonable request.

\appendix
\section{\label{appendixa}Comparison of numerical performances between MPS and TTNS}
In this appendix, we demonstrate the advantages of using the TTNS approach over the MPS/TT for studying chemical reactions inside an optical cavity. MPS/TT represents a special, simplified form of TTNS, where all nodes are arranged in a one-dimensional chain, as illustrated in \Fig{MPSvsTTNS}. Although MPS/TT restricts the network topology, its numerical performance is still heavily influenced by the ordering of physical indices.
\Fig{MPSvsTTNS}b and \Fig{MPSvsTTNS}c compare $k(t)$ obtained with two different index orderings in the MPS. In the first configuration, molecular nodes and their associated solvent bath nodes are attached subsequently to the cavity and cavity bath nodes on the right. While each molecule couples equally to the cavity mode, the higher-index molecular nodes are positioned farther from the cavity node, giving rise to long-distance entanglement and thus necessitating a considerably large bond dimension to accurately capture the system dynamics. This issue can be partially mitigated by arranging half of the molecular and associated bath nodes on either side of the cavity nodes, as shown in \Fig{MPSvsTTNS}c. Nevertheless, a large bond dimension is still required to obtain smooth and converged results. 

In the TTNS decomposition scheme tailored to the collective regime of vibrational polariton chemistry, each molecular node is positioned equidistant from the cavity node, separated by only a few connecting nodes. This tree topology and index arrangement significantly reduce the need for large bond dimensions, optimizing computational efficiency while preserving accuracy. For instance, simulations with a bond dimension of $D_{\rm max}=20$ are sufficient to produce high-quality results, as shown in \Fig{MPSvsTTNS}d.

%

\pagebreak
\widetext
\clearpage
\begin{center}
\textbf{\large Supplementary information: Stochastic resonance in vibrational polariton chemistry}
\end{center}
\setcounter{equation}{0}
\setcounter{figure}{0}
\setcounter{table}{0}
\setcounter{page}{1}
\makeatletter
\renewcommand{\theequation}{S\arabic{equation}}
\renewcommand{\thefigure}{S\arabic{figure}}
\renewcommand{\bibnumfmt}[1]{[S#1]}
\renewcommand{\citenumfont}[1]{S#1}
\renewcommand{\Sec}[1]{Sec.\,\ref{#1}}

\section*{Impact of cavity bath parameters $\lambda_{\mathrm{c}}$ and $\Omega_{\mathrm{c}}$ on the damping of system dynamics}
\begin{figure}[h]
    \centering
    \begin{minipage}[c]{0.45\textwidth}
    \raggedright a) $\lambda_{\mathrm{c}}=50\,\text{cm}^{-1}$
        \includegraphics[width=\textwidth]{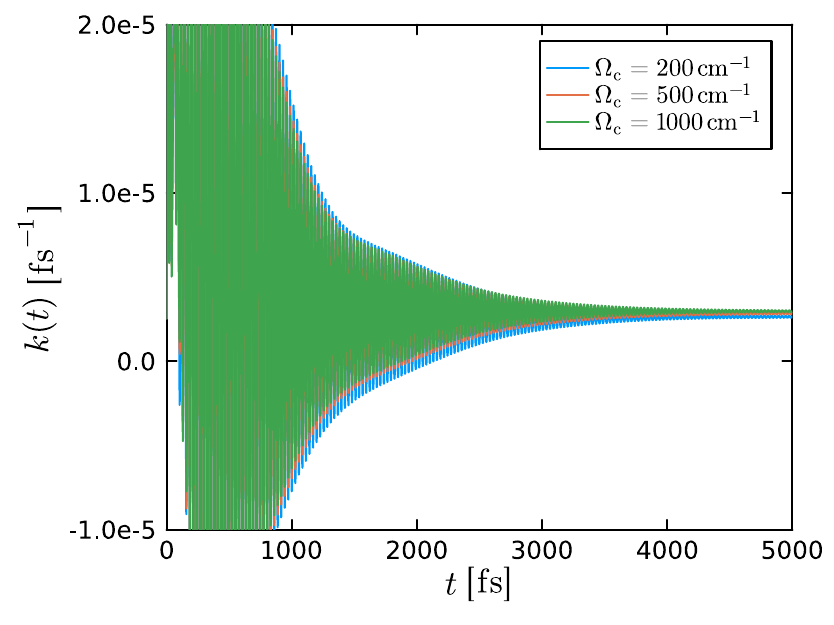}
    \end{minipage}
    \begin{minipage}[c]{0.45\textwidth}
    \raggedright b) $\lambda_{\mathrm{c}}=50\,\text{cm}^{-1}$
        \includegraphics[width=\textwidth]{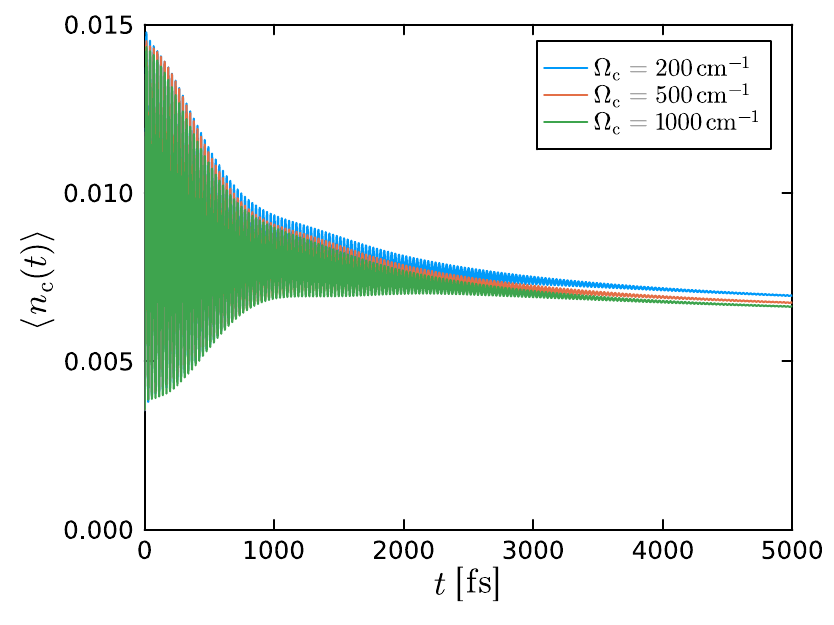}
    \end{minipage}
    \begin{minipage}[c]{0.45\textwidth}
    \raggedright c) $\Omega_{\mathrm{c}}=1000\,\text{cm}^{-1}$
        \includegraphics[width=\textwidth]{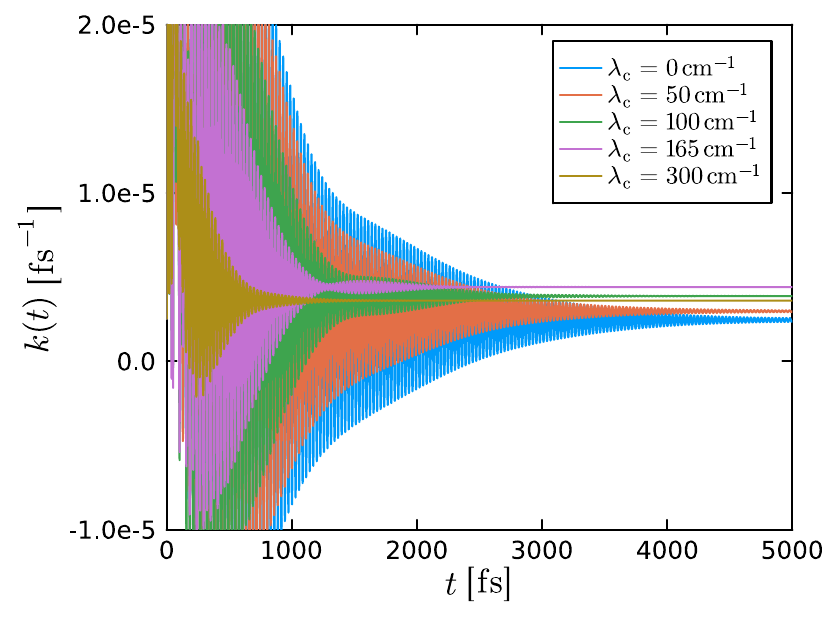}
    \end{minipage}
    \begin{minipage}[c]{0.45\textwidth}
    \raggedright d) $\Omega_{\mathrm{c}}=1000\,\text{cm}^{-1}$
        \includegraphics[width=\textwidth]{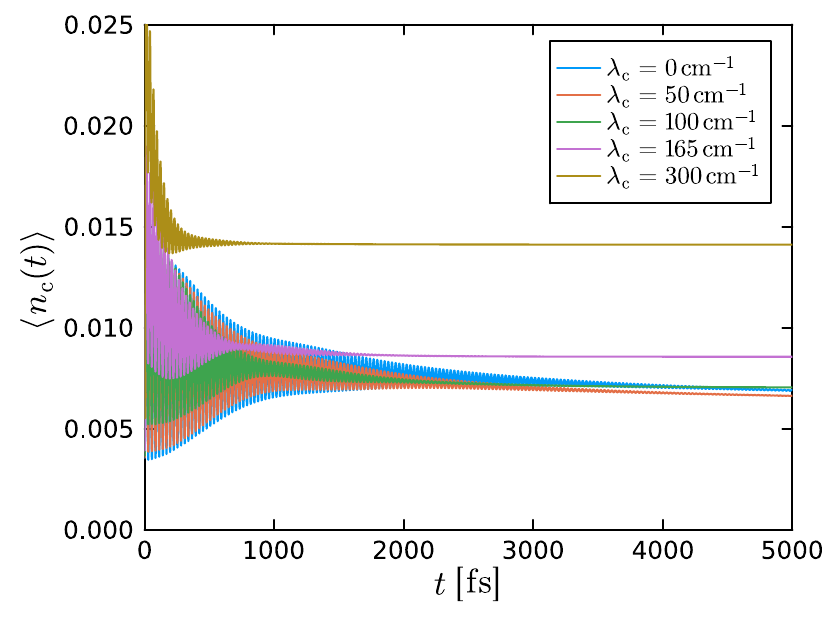}
    \end{minipage}
\caption{\label{bathparameters}
Rescaled flux-side correlation function $k(t)$, and the dynamics of the average photonic excitation number $\langle n_{\mathrm{c}}(t)\rangle$, are illustrated for different values of $\Omega_{\mathrm{c}}$ with $\lambda_{\mathrm{c}}=50\,\text{cm}^{-1}$ in (a) and (b), respectively. Similarly, the behavior for varying $\lambda_{\mathrm{c}}$ with a fixed $\Omega_{\mathrm{c}}=1000\,\text{cm}^{-1}$ is shown in (c) and (d). Other parameters are set as follows: $\omega_{\mathrm{c}}=1185\,\text{cm}^{-1}$, $\eta_{\mathrm{c}}=0.00125$\au, $T=300\,\text{K}$, $d_{\mathrm{c}}=10$, and $N_{\mathrm{mol}}=1$.}
\end{figure}

To elucidate the influence of cavity bath parameters on system dynamics, \Fig{bathparameters} presents the rescaled flux-side correlation function $k(t)$, and the time evolution of the average photonic excitation number $\langle n_{\mathrm{c}}(t)\rangle$, for various values of $\lambda_{\mathrm{c}}$ and $\Omega_{\mathrm{c}}$. In \Fig{bathparameters}a and \Fig{bathparameters}b, $\Omega_{\mathrm{c}}$ is varied while $\lambda_{\mathrm{c}}$ is held constant at $50\,\text{cm}^{-1}$. Under this condition, variations in $\Omega_{\mathrm{c}}$ do not significantly affect the damping rates of either the molecular vibrational or cavity photonic dynamics. In comparison, \Fig{bathparameters}c and \Fig{bathparameters}d depict results for varying $\lambda_{\mathrm{c}}$ with $\Omega_{\mathrm{c}}=1000\,\text{cm}^{-1}$, where both $k(t)$ and $\langle n_{\mathrm{c}}(t)\rangle$ exhibit accelerated damping as $\lambda_{\mathrm{c}}$ increases. These findings justify the use of parameter $\lambda_{\mathrm{c}}$ in this work as a quantifier of the cavity damping strength. 

\section*{Impact of bath characteristic frequency $\Omega_{\alpha}$ on reaction dynamics}
\begin{figure}
    \centering
    \begin{minipage}[c]{0.45\textwidth}
        \raggedright a) 
        \includegraphics[width=\textwidth]{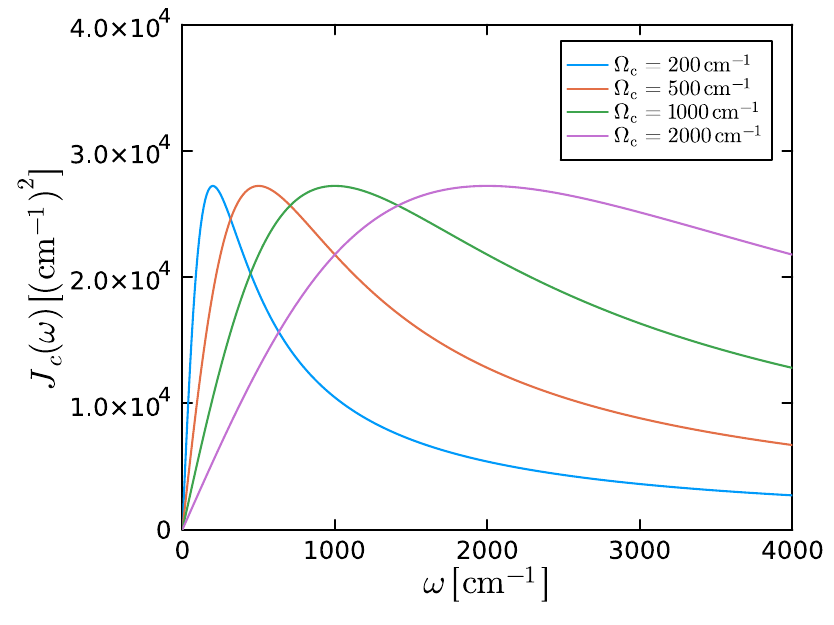}
    \end{minipage}
    \begin{minipage}[c]{0.45\textwidth}
        \raggedright b)
        \includegraphics[width=\textwidth]{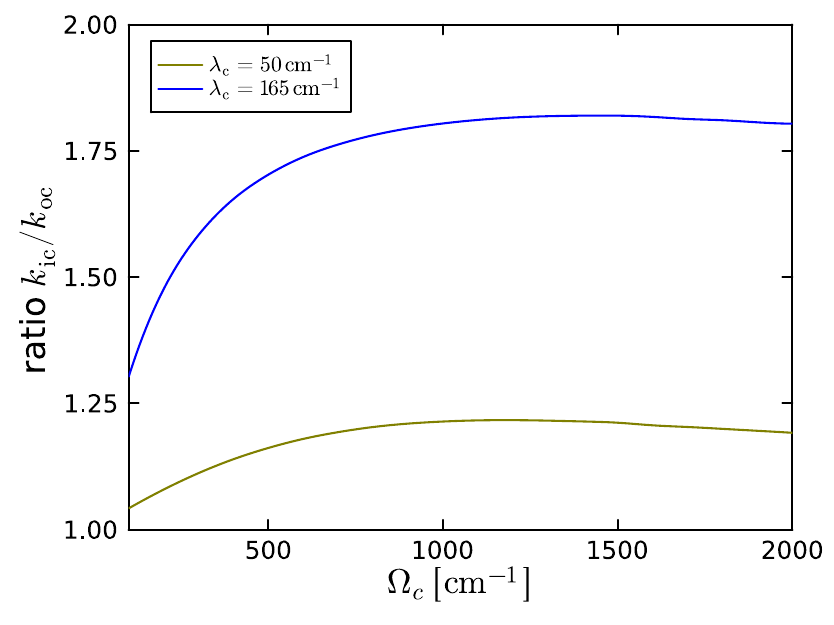}
    \end{minipage}
    \begin{minipage}[c]{0.45\textwidth}
        \raggedright c) 
        \includegraphics[width=\textwidth]{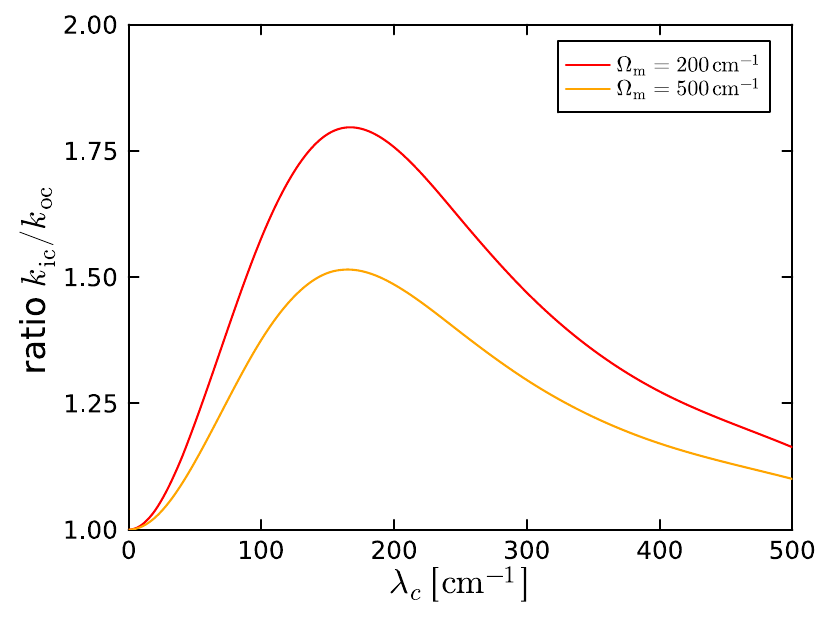}
    \end{minipage}
\caption{\label{bandwidth}
a) Spectral density function of the cavity bath is shown for $\lambda_{\mathrm{c}}=100\,\text{cm}^{-1}$ with three different values of the characteristic frequency $\Omega_{\mathrm{c}}$, highlighting how $\Omega_{\mathrm{c}}$ influences the peak and width of the function. b) Ratio of rate $k_{\mathrm{ic}}/k_{\mathrm{oc}}$ as a function of $\Omega_{\mathrm{c}}$ for two distinct values of $\lambda_{\mathrm{c}}$. These simulations use  $P=6$ Pad\'e poles. c) Ratio of rate $k_{\mathrm{ic}}/k_{\mathrm{oc}}$ as a function of $\lambda_{\mathrm{c}}$ for a fixed $\Omega_{\mathrm{c}}=1000\,\text{cm}^{-1}$ and two different values of the solvent bath characteristic frequency, $\Omega_{\mathrm{m}}$.  Other parameters used in the simulations are $\omega_c=1185\,\text{cm}^{-1}$, $\eta_{\mathrm{c}}=0.00125$\au, $T=300\,\text{K}$, $d_{\mathrm{c}}=10$, and $N_{\mathrm{mol}}=1$.}
\end{figure}

In this work, the spectral density functions of both the cavity bath and solvent baths are modeled using a Debye-Lorentzian form, characterized by the parameters $\lambda_{\alpha}$ and $\Omega_{\alpha}$. The characteristic frequency $\Omega_{\alpha}$ defines both the peak position and the width of the spectral density function, with a maximum value of $J(\Omega_{\alpha})=\lambda_{\alpha}^2$, as depicted in \Fig{bandwidth}a.

For the cavity bath, increasing $\Omega_{\mathrm{c}}$ initially enhances the reaction rate inside the cavity, followed by a plateau and a gradual decrease at higher $\Omega_{\mathrm{c}}$ values, as shown in \Fig{bandwidth}b. Moreover, the reaction dynamics are also influenced by the characteristic frequency of the solvent bath,  $\Omega_m$. A larger $\Omega_{\mathrm{m}}$ increases the reaction rate outside the cavity. For example, the rate $k_{\mathrm{oc}}$ rises from $2.46\times10^{-6} \,\text{fs}^{-1}$  at $\Omega_{\mathrm{m}}=200\,\text{cm}^{-1}$ to $3.8\,\times 10^{-6} \,\text{fs}^{-1}$ at $\Omega_{\mathrm{m}}=500\,\text{cm}^{-1}$. In contrast, a larger $\Omega_{\mathrm{m}}$ results in less pronounced changes in the rates inside the cavity, while preserving the stochastic resonance feature, i.e. the turnover of $k_{\mathrm{ic}}/k_{\mathrm{oc}}$ versus $\lambda_{\mathrm{c}}$, as illustrated in \Fig{bandwidth}c. These findings highlight the differential effects of $\Omega_{\alpha}$ on reaction rates, depending on the bath's nature.

\section*{Impact of the light-matter coupling strength on reaction dynamics}
\begin{figure}
    \centering
    \begin{minipage}[c]{0.5\textwidth}
        \includegraphics[width=\textwidth]{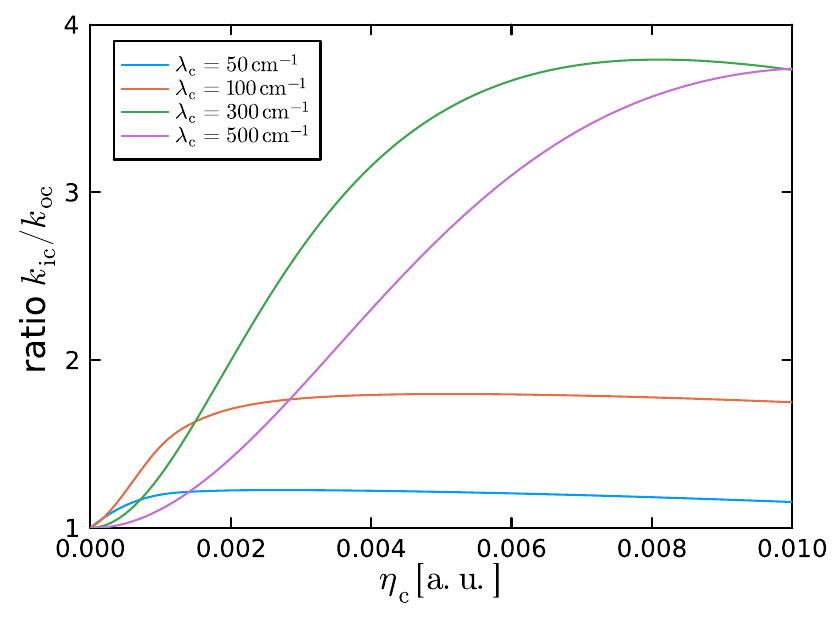}
    \end{minipage}
\caption{\label{etac} Ratio of rates $k_{\mathrm{ic}}/k_{\mathrm{oc}}$ as a function of the light--matter coupling strength $\eta_{\mathrm{c}}$ in the single molecule limit for varying values of $\lambda_{\mathrm{c}}$.  Other parameters used in the simulations are $\omega_{\mathrm{c}}=1185\,\text{cm}^{-1}$, $\Omega_{\mathrm{c}}=1000\,\text{cm}^{-1}$,  $T=300,\text{K}$, and $d_{\mathrm{c}}=10$.}
\end{figure}

In the main text, it is observed that increasing the light-matter coupling strength, $\eta_{\mathrm{c}}$, shifts the peak position in the $k_{\mathrm{ic}}/k_{\mathrm{oc}}$ versus $\lambda_{\mathrm{c}}$ profile to a larger value and simultaneously increases the peak height. To systematically demonstrate the impact of light-matter coupling strength, \Fig{etac} presents the cavity-induced rate modification $k_{\mathrm{ic}}/k_{\mathrm{oc}}$ as a function of $\eta_{\mathrm{c}}$ for varying cavity damping strengths. The cavity frequency is fixed at $\omega_{\mathrm{c}}=1185\,\mathrm{cm}^{-1}$, near resonance with vibrational transitions, and other parameters are $T=300$K and $\Omega_{\mathrm{c}}=1000\,\mathrm{cm}^{-1}$. Interestingly, the reaction rates exhibit a turnover behavior as $\eta_{\mathrm{c}}$ increases. In a weakly damped cavity $(\lambda_c=50\,\mathrm{cm}^{-1})$, the maximum rate ratio is $k_{\mathrm{ic}}/k_{\mathrm{oc}}=1.225$, occurring at $\eta_{\mathrm{c}}=0.0025$\au  At higher noise levels (larger $\lambda_{\mathrm{c}}$), while the rate enhancement inside the cavity is initially suppressed in the small light-matter coupling regime, it increases rapidly with stronger light-matter coupling strength. The maximum rate at higher $\lambda_{\mathrm{c}}$ is significantly greater than that at lower $\lambda_{\mathrm{c}}$. For instance, at $\lambda_{\mathrm{c}}=300\,\mathrm{cm}^{-1}$, the rate inside the cavity can be enhanced by a factor of 3.8 with a strong light-matter coupling strength $\eta_c=0.008$\au 

\section*{Infrared absorption line shapes and rate modification profiles in the single-molecule limit}
\begin{figure}
    \centering
    \begin{minipage}[c]{0.33\textwidth}
        \raggedright a) $\Omega_c=1000\,\text{cm}^{-1}$
        \includegraphics[width=\textwidth]{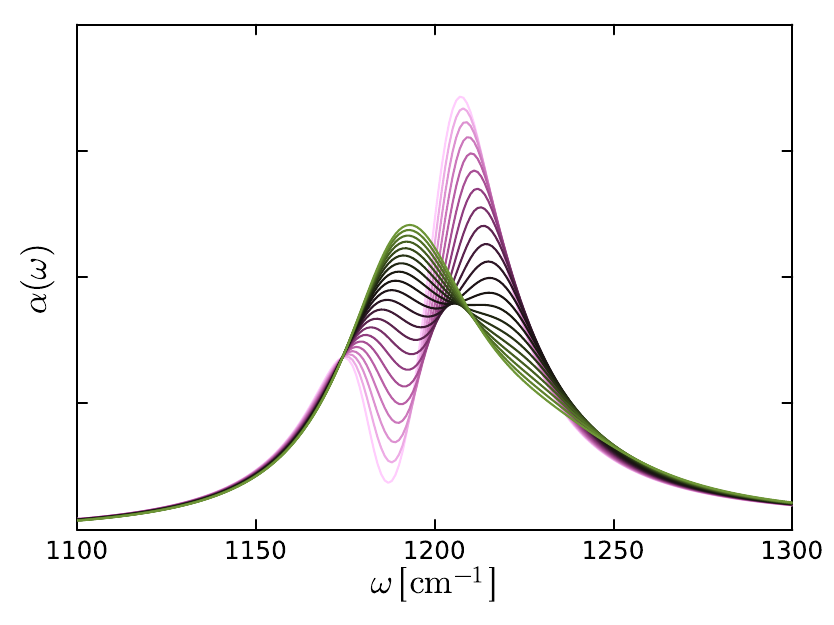}
    \end{minipage}
        \begin{minipage}[c]{0.33\textwidth}
        \raggedright b) $\Omega_c=500\,\text{cm}^{-1}$
        \includegraphics[width=\textwidth]{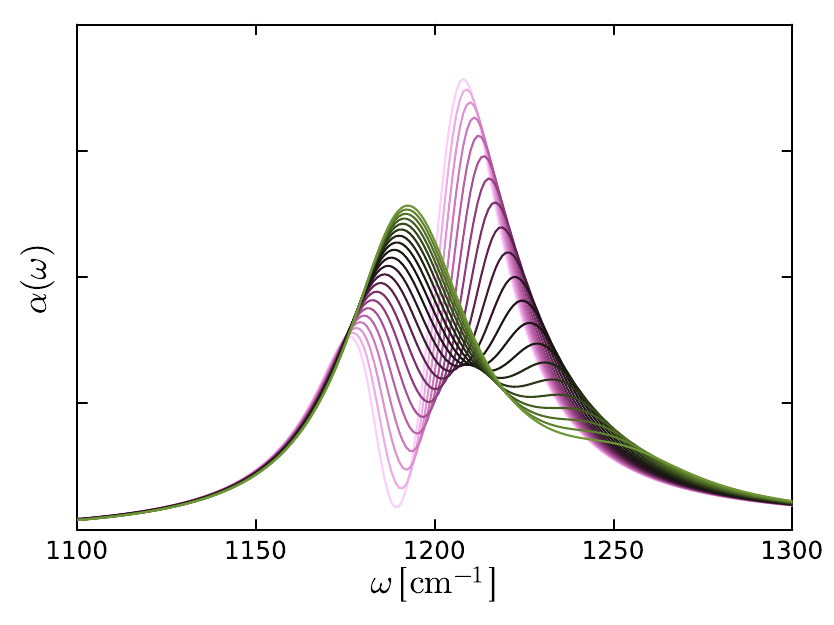}
    \end{minipage}
        \begin{minipage}[c]{0.33\textwidth}
        \raggedright c) $\Omega_c=200\,\text{cm}^{-1}$
        \includegraphics[width=\textwidth]{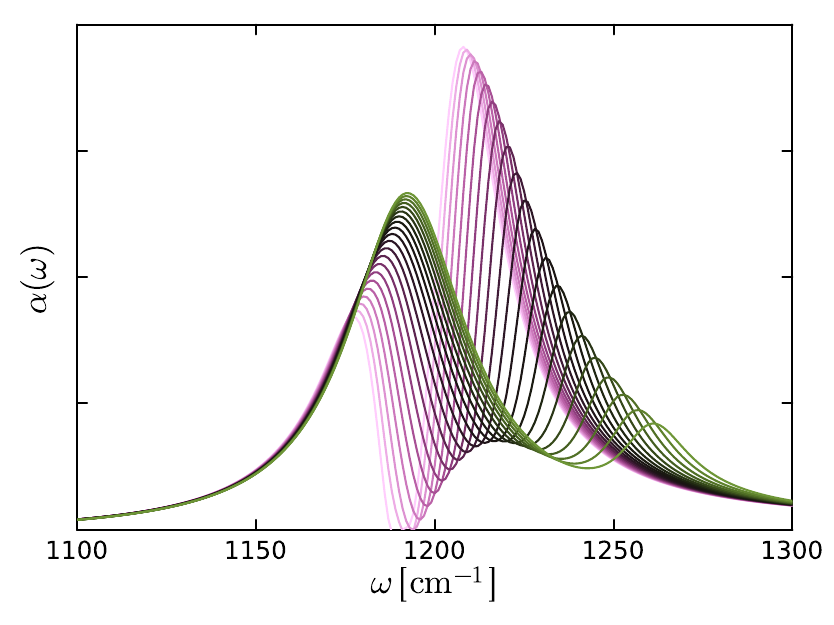}
    \end{minipage}
\caption{\label{absorption}
Infrared absorption line shapes of a single molecule ($N_{\mathrm{mol}}=1$) inside the cavity for $\lambda_{\mathrm{c}}$ values ranging from $100\,\text{cm}^{-1}$ (pink) to $300\,\text{cm}^{-1}$(green), with increments of $10\,\text{cm}^{-1}$. Each panel represents the spectra for a different cavity bath characteristic frequency $\Omega_{\mathrm{c}}$. The other simulation parameters are as follows: $\omega_c=1185\,\text{cm}^{-1}$, $\eta_{\mathrm{c}}=0.00125$\au, $T=300\,\text{K}$, and $d_{\mathrm{c}}=10$.}
\end{figure}
\begin{figure}
    \centering
    \begin{minipage}[c]{0.5\textwidth}
        \includegraphics[width=\textwidth]{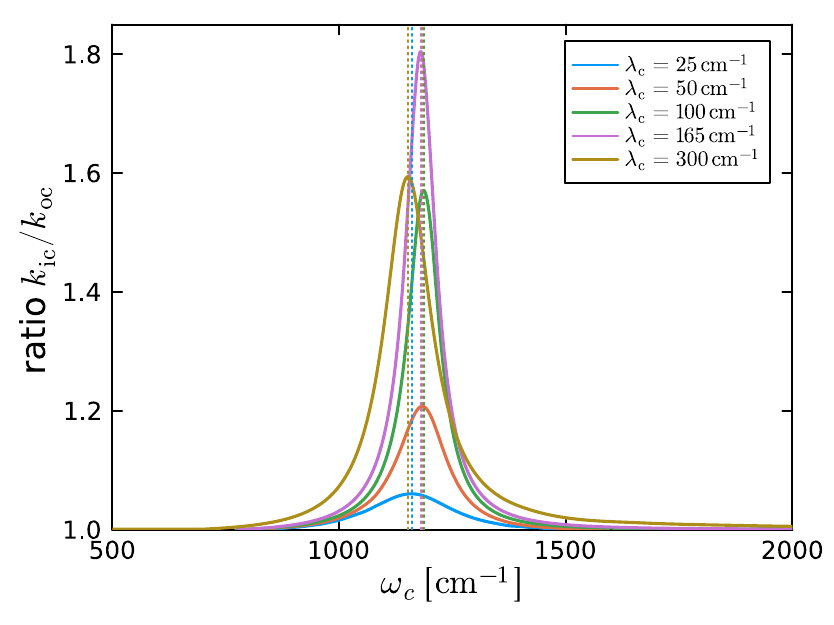}
    \end{minipage}
\caption{\label{ratetodamping}
 Rate modification profile as a function of the cavity frequency in the single-molecule limit for different values of the cavity damping strength, $\lambda_c$. The vertical dotted lines in the figure mark the position of the peaks. Other parameters used in the simulations are $\Omega_{\rm c}=1000\,\text{cm}^{-1}$, $\eta_{\rm c}=0.00125$\au, $T=300\,\text{K}$, and $d_{\rm c}=10$.  }
\end{figure}

\Fig{absorption} illustrates the absorption line shapes, $\alpha(\omega)\propto \frac{1}{2\pi}\int_{-\infty}^{\infty}\mathrm{d}t e^{i\omega t}\langle \vec{\mu}(x,t)\vec{\mu}(x,0)\rangle$, which is obtained by performing the Fourier transform of the dipole-dipole auto-correlation function. These absorption spectra are calculated for cavity damping strengths $\lambda_{\mathrm{c}}$ ranging from $100\,\text{cm}^{-1}$ to $300\,\text{cm}^{-1}$, covering the turnover observed in Fig.2 of the main text, and for three different values of the cavity bath's characteristic frequency $\Omega_{\mathrm{c}}$ in the single-molecule limit. 

The characteristic frequency of the cavity bath $\Omega_{\mathrm{c}}$ is inversely related to the bath's characteristic timescale, with smaller $\Omega_{\mathrm{c}}$ corresponding to slower bath response and more pronounced non-Markovian feature. In the Markovian limit $(\Omega_{\mathrm{c}}=1000\,\text{cm}^{-1})$, split double peaks corresponding to the upper and lower polaritons are observed for $\lambda_{\mathrm{c}}<250\,\text{cm}^{-1}$. These peaks broaden and blue shift as $\lambda_{\mathrm{c}}$ increases, although the Rabi splitting, i.e. the energy gap between two peaks, remains nearly constant, $\Omega_R=32\,\text{cm}^{-1}$. In contrast, in the non-Markovian regime $(\Omega_{\mathrm{c}}=200\,\text{cm}^{-1})$, the Rabi splitting remains evident event at higher $\lambda_{\mathrm{c}}$ and increases from $\Omega_R=32\,\text{cm}^{-1}$ for $\lambda_{\rm c}=100\,\text{cm}^{-1}$ to $\Omega_R=68\,\text{cm}^{-1}$ for $\lambda_{\rm c}=300\,\text{cm}^{-1}$.

In addition, the rate modification profiles as a function of the cavity frequency $\omega_{\mathrm{c}}$ are shown in \Fig{ratetodamping} for various $\lambda_{\mathrm{c}}$ with a fixed characteristic frequency of the cavity bath, $\Omega_{\mathrm{c}}=1000\,\text{cm}^{-1}$. As $\lambda_{\mathrm{c}}$ increases, the resonant peak first shifts to larger $\omega_{\mathrm{c}}$ before reversing direction.  This shift arises from the broadening effects of the molecular vibrational and cavity photonic states due to their coupling with the baths. For the double-well model considered in this work, two vibrational transitions with the energies
$1140\,\text{cm}^{-1}$ and $1238\,\text{cm}^{-1}$ are involved in the reaction mechanisms. Photons with frequencies between these two vibrational transitions
can trigger either transition, enabling an additional cavity-induced intramolecular reaction pathway and then altering the reaction rate, as analyzed in our previous work.\cite{Ke_J.Chem.Phys._2024_p224704} The peak position also varies with changes in the molecule–solvent coupling strength ($\lambda_{i}$), the light-matter coupling strength $\eta_{\mathrm{c}}$.\cite{Ke_2024_JCP_p54104} Furthermore, as the system approaches stochastic resonance, i.e. the optimal intermediate noise level at $\lambda_{\mathrm{c}}=165\,\text{cm}^{-1}$, the peak becomes sharper and stronger, before broadening and weakening with further increase in the cavity damping strength $\lambda_{\mathrm{c}}$. 

The peak positions and full width at half maximum (FWHM) values for different $\lambda_{\mathrm{c}}$ are summarized below: 
\begin{center}
\begin{tabular}{ c | c | c | c | c | c} 
 $\lambda_{\mathrm{c}}\,[\text{cm}^{-1}]$ & $25$ & $50$ & $100$ & $165$ & $300$ \\
 \hline
Peak position $[\text{cm}^{-1}]$ & $1160$  &  $1184$ & $1187$ & $1180$  & $1152$ \\  
\hline
FWHM $[\text{cm}^{-1}]$ & $205$ & $125$ & $85$ & $80$ &   $125$  
\end{tabular}
\end{center}

Interestingly, a sharp resonant peak in the rate modification profile can be observed regardless of whether the Rabi splitting is present (e.g. when $\lambda_{\mathrm{c}}=100\,\text{cm}^{-1}$) or absent (e.g. when $\lambda_{\mathrm{c}}=300\,\text{cm}^{-1}$). In the vibrational strong coupling regime, characterized by Rabi splitting significantly exceeding the FWHM of the split peaks in the absorption line shapes, e.g. when $\lambda_{\mathrm{c}}< 50\,\text{cm}^{-1}$, only a broad and weak resonant peak is observed. These results suggest that observing a Rabi splitting is neither a necessary nor a sufficient condition for the emergence of pronounced resonant rate modification in vibrational polariton chemistry.

\section*{Importance of higher photonic excitation manifold}
\begin{figure}
    \centering
    \begin{minipage}[c]{0.45\textwidth}
    \raggedright a) 
        \includegraphics[width=\textwidth]{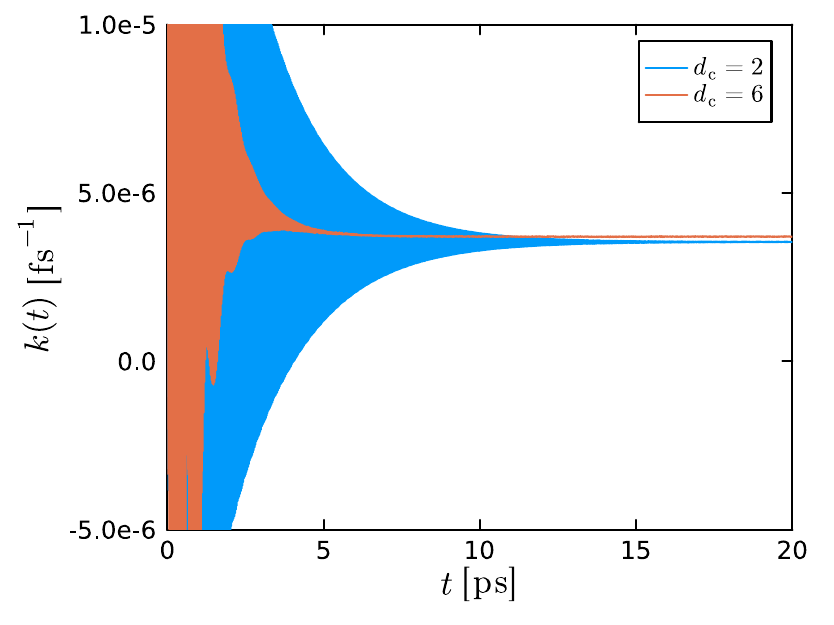}
    \end{minipage}
    \begin{minipage}[c]{0.45\textwidth}
    \raggedright b) 
        \includegraphics[width=\textwidth]{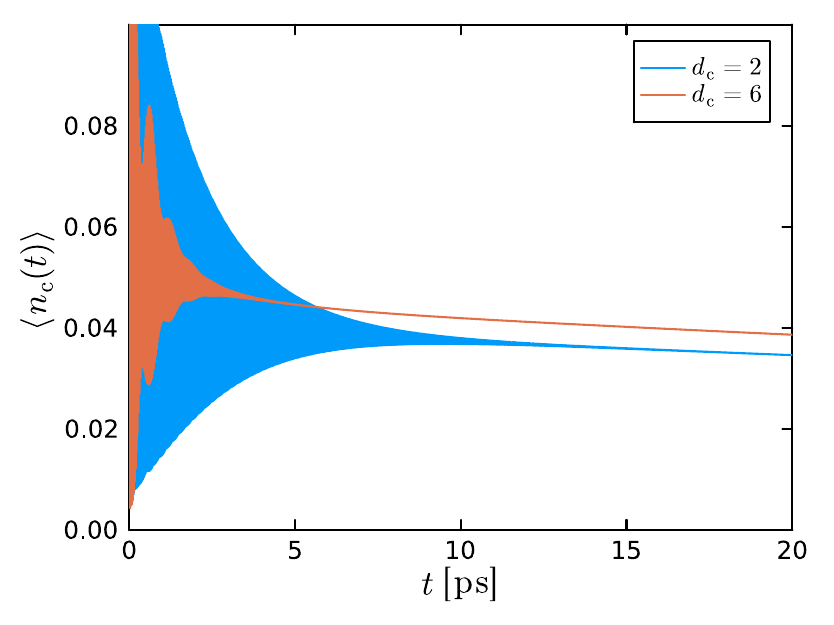}
    \end{minipage}
\caption{\label{multiphoton_collective}
Rescaled flux-side correlation function $k(t)$ (panel a) and the dynamics of the average photonic excitation number $\langle n_{\mathrm{c}}(t)\rangle$ (panel b) for simulations including two ($d_{\mathrm{c}}=2$) and ten ($d_{\mathrm{c}}=10$) lowest photonic states. Other parameters used in the simulations are $\lambda_{\mathrm{c}}=50\,\text{cm}^{-1}$, $\Omega_{\mathrm{c}}=1000\,\text{cm}^{-1}$, $\omega_c=1185\,\text{cm}^{-1}$, $\eta_{\mathrm{c}}=0.00125$\au, $T=300\,\text{K}$, and $N_{\mathrm{mol}}=4$. These results underscore the necessity of including higher photonic states for accurate modeling in the collective coupling regime.}
\end{figure}
\begin{figure}
    \centering
    \begin{minipage}[c]{0.45\textwidth}
        \raggedright a)  
        \includegraphics[width=\textwidth]{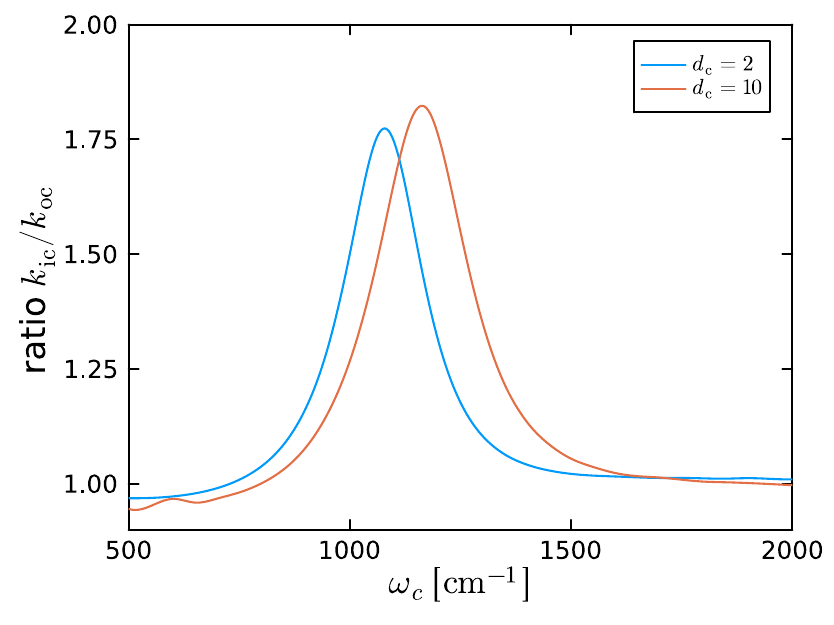}
    \end{minipage}
    \begin{minipage}[c]{0.45\textwidth}
        \raggedright b) 
        \includegraphics[width=\textwidth]{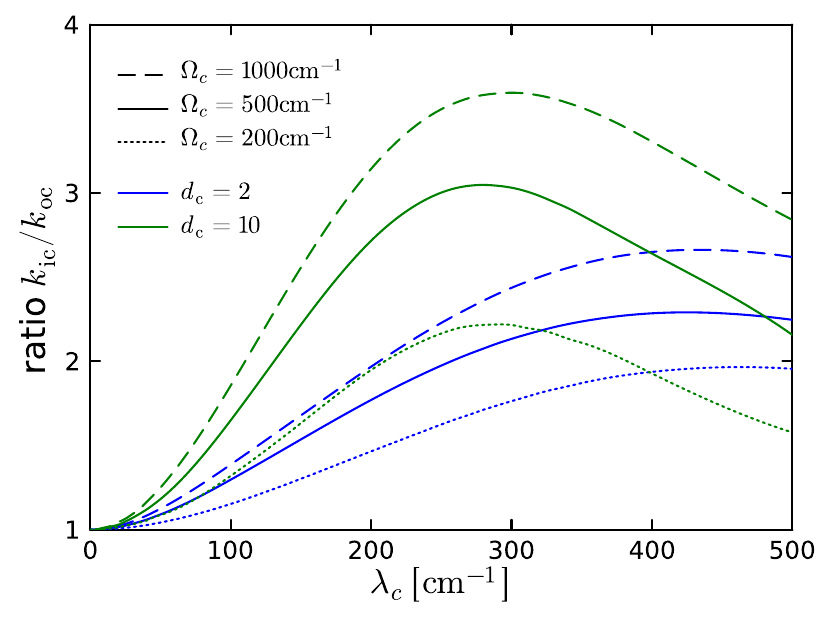}
    \end{minipage}
\caption{\label{multiphoton_singlemolecule}
a) Rate modification profile $k_{\mathrm{ic}}/k_{\mathrm{oc}}$ as a function of the cavity frequency $\omega_c$ for single-molecule simulations including two ($d_{\mathrm{c}}=2$) and ten ($d_{\mathrm{c}}=10$) lowest photonic excited states, respectively. The parameters for the cavity bath are $\lambda_{\mathrm{c}}=100\,\text{cm}^{-1}$ and $\Omega_{\mathrm{c}}=1000\,\text{cm}^{-1}$. The light-matter coupling strength is $\eta_{\mathrm{c}}=0.005\,$a.u., which results in a Rabi splitting of $\Omega_R=108\, \text{cm}^{-1}$ under the resonant condition. b) Ratio of rate $k_{\mathrm{ic}}/k_{\mathrm{oc}}$ as a function of $\lambda_{\mathrm{c}}$ for three different cavity bath characteristic frequencies $\Omega_{\mathrm{c}}$ under resonant condition ($\omega_{\rm c}=1185\,\text{cm}^{-1}$), strong light-matter coupling ($\eta_{\rm c}=0.005$\au), single-molecule ($N_{\mathrm{mol}}=1$) limit. The blue and green lines correspond to simulations with $d_{\mathrm{c}}=2$ and  $d_{\mathrm{c}}=10$, respectively. The temperature is set at $T=300\,\text{K}$. These results highlight the importance of multi-photon processes in accurate rate predictions in the strong light-matter coupling regime.}
\end{figure}

In the main text, it has been demonstrated that accounting for higher photonic excitation states is essential for accurately capturing reaction dynamics, particularly in the collective coupling regime.  When only the lowest photonic excited state
($d_{\mathrm{c}}=2$) is considered, both the reaction rates and the damping of
oscillations in the rescaled flux-side correlation function $k(t)$, and the average photonic excitation number $\langle n_{\mathrm{c}}(t)\rangle$, are significantly underestimated. This is illustrated in \Fig{multiphoton_collective} for the case of $N_{\mathrm{mol}}=4$.

Furthermore, even in the single-molecule limit, the inclusion of higher photonic excitation states is critical. \Fig{multiphoton_singlemolecule} compares the results for the light-matter coupling strength $\eta_{\mathrm{c}}=0.005\,$\au with $d_{\mathrm{c}}=2$ and $d_{\mathrm{c}}=10$, the latter representing the convergent solution. Neglecting multi-photon processes leads not only to incorrect predictions of the resonant peak in the rate modification profile as a function of the cavity frequency $\omega_{\mathrm{c}}$, but also to an overestimation of the turnover $\lambda_{\mathrm{c}}$ in the stochastic resonance effect.

These findings underscore the importance of considering the higher photonic excitation manifold to accurately capture the interplay between light-matter coupling and vibrational dynamics in both single-molecule and collective scenarios.

\section*{Impact of various noise levels in collective coupling regime}
 \begin{figure}
    \centering
    \begin{minipage}[c]{0.5\textwidth}
        \includegraphics[width=\textwidth]{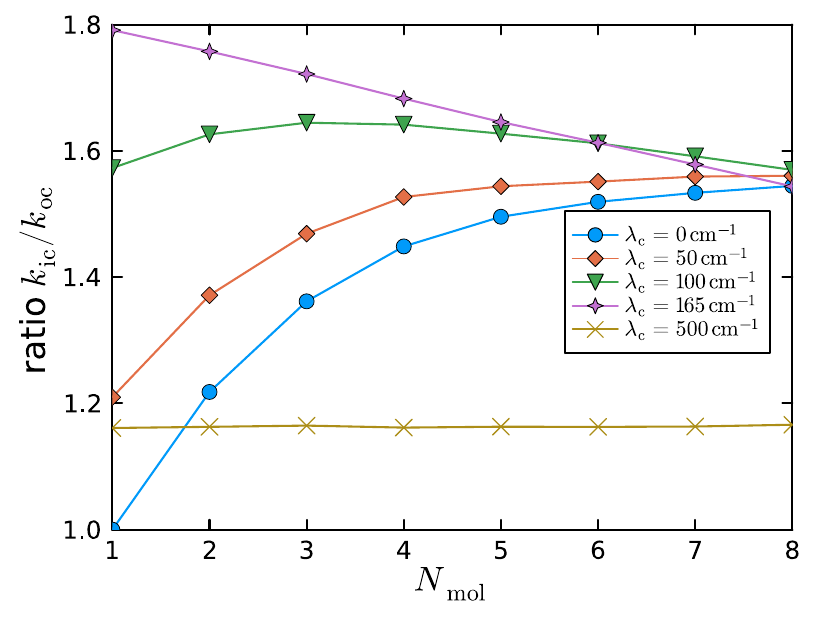}
    \end{minipage}
\caption{\label{noiselevel} Ratio of rates $k_{\mathrm{ic}}/k_{\mathrm{oc}}$ as a function of the number of molecules collectively coupled to the cavity mode for varying values of $\lambda_{\mathrm{c}}$.  Other parameters used in the simulations are $\omega_{\mathrm{c}}=1185\,\text{cm}^{-1}$, $\Omega_{\mathrm{c}}=1000\,\text{cm}^{-1}$, $\eta_{\mathrm{c}}=0.00125\,$\au, $T=300,\text{K}$, and $d_{\mathrm{c}}=8$.}
\end{figure}

\Fig{noiselevel} displays the ratio of rates $k_{\mathrm{ic}}/k_{\mathrm{oc}}$ as a function of $N_{\mathrm{mol}}$ for varing levels of noise ($\lambda_{\mathrm{c}}$). In a lossless cavity ($\lambda_{\mathrm{c}}=0\,\text{cm}^{-1}$) and a weak cavity damping scenario ($\lambda_{\mathrm{c}}=50\,\text{cm}^{-1}$), reaction rates inside the cavity exhibit a rapid increase as the number of molecules aggregates. At an intermediate noise level ($\lambda_{\mathrm{c}}=100\,\text{cm}^{-1}$), the rate enhancement initially rises with $N_{\mathrm{mol}}$ but then declines as the number of molecules continues to grow. For a stronger cavity damping ($\lambda_{\mathrm{c}}=165\,\text{cm}^{-1}$), the rate enhancement steadily diminishes with increasing $N_{\mathrm{mol}}$. However, in the case of ultrastrong damping ($\lambda_{\mathrm{c}}=300\,\text{cm}^{-1}$), while the rate enhancement in the single-molecule limit is not substantial, it declines very slowly with the addition of more connected molecules.

These results highlight the strong dependence of stochastic resonance on system size, i.e. the number of molecules in a molecular network connected by the cavity mode. As seen in \Fig{noiselevel}, the sensitivity of the rate modification to $\lambda_{\mathrm{c}}$ is reduced in larger systems, with a notable rate enhancement even in the weak cavity damping regime. This behavior is likely attributed to the rapid increase in the number of cavity-induced intermolecular reaction pathways,\cite{Ke_J.Chem.Phys._2024_p224704} which grows proportionally to $N_{\mathrm{mol}}\times (N_{\mathrm{mol}}-1)/2$,  binding all molecules together in the collective process. Further studies will delve deeper into the relationship of the reactivity and the degree of intermolecular delocalization in large systems.  

\section*{Impact of dipole orientations}
 \begin{figure}
    \centering
    \begin{minipage}[c]{0.45\textwidth}
    \raggedright a) 
        \includegraphics[width=\textwidth]{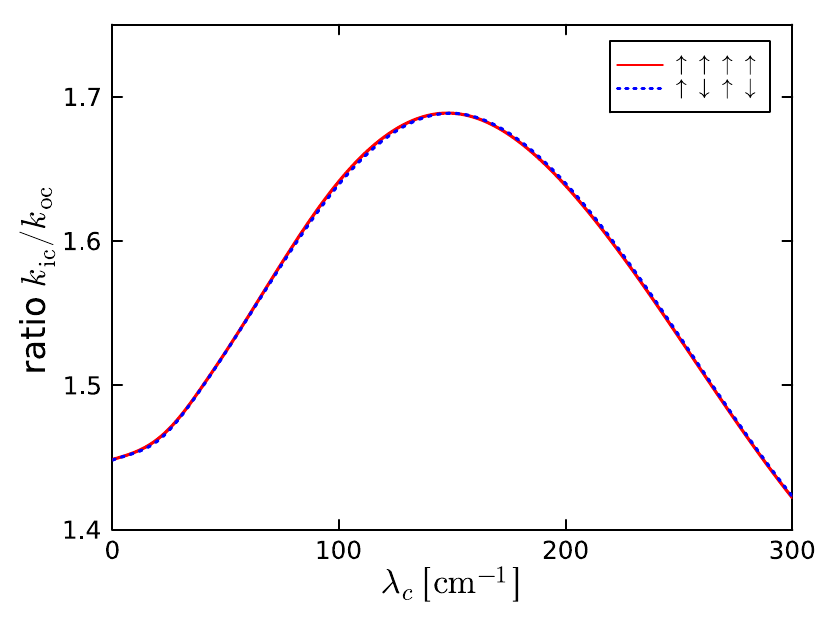}
    \end{minipage}
    \begin{minipage}[c]{0.45\textwidth}
    \raggedright b) 
        \includegraphics[width=\textwidth]{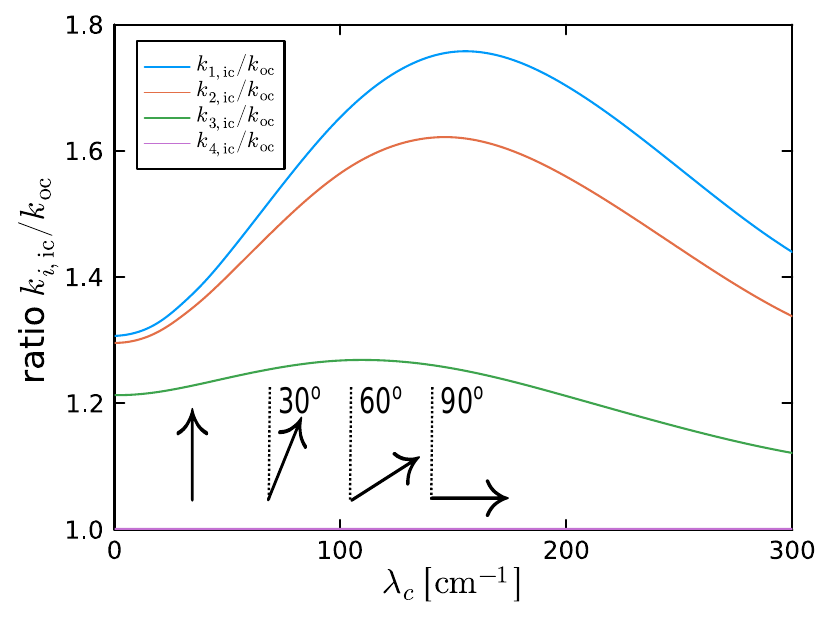}
    \end{minipage}
\caption{\label{dipoleorientation} a) Ratio of rates $k_{\mathrm{ic}}/k_{\mathrm{oc}}$ as a function of the cavity damping strength $\lambda_c$ for $N_{\mathrm{mol}}=4$. The red solid line represents the aligned dipole configuration, where all molecular dipoles are aligned with the light polarization ($\theta_i=0$). The blue dotted line corresponds to the configuration where the molecular dipoles of the first and third molecules align with the light polarization ($\theta_{1/3}=0$), while those of the second and fourth molecules point in the opposite direction ($\theta_{2/4}=\pi$). b) Ratio of rates $k_{\mathrm{i, ic}}/k_{\mathrm{oc}}$ for individual molecules as a function of $\lambda_c$ in a misaligned molecular ensemble with $N_{\mathrm{mol}}=4$. The dipole angles relative to the light polarization are $\theta_i=(i-1)\pi/6$, leading to progressively weaker effective light-matter coupling strength for molecules further misaligned from the light polarization axis. The first molecule $(\theta_1=0)$ achieves the maximum coupling strength, exhibiting larger rate enhancements and higher turnovers in $\lambda_{\mathrm{c}}$, while the fourth molecule $(\theta_4=\pi/2)$ is fully decoupled from the cavity mode, showing no rate modification across the entire $\lambda_{\mathrm{c}}$ range. Other parameters used in the simulations are $\omega_{\mathrm{c}}=1185\,\text{cm}^{-1}$, $\Omega_{\mathrm{c}}=1000\,\text{cm}^{-1}$, $\eta_{\mathrm{c}}=0.00125\,$\au, $T=300\,\text{K}$, and $d_{\mathrm{c}}=8$. }
\end{figure}

In our prior study,\cite{Ke_J.Chem.Phys._2024_p224704}  we have demonstrated that for a dimer, as long as the molecular dipoles are in line with the light polarization ($\cos\theta_i = \pm 1$ in $\vec{\mu}_i(x_i)\cdot \vec{e}=x_i\cos\theta_i$), the reaction rates remains unchanged regardless of whether $\theta_i=0$ or $\theta_i=\pi$, although the transient dynamics are significantly influenced by the molecular dipole orientation. This phenomenon persists even with more molecules. Specifically, reaction rates are invariant to the sign in $\cos\theta_i$. An example with $N_{\mathrm{mol}}=4$ and two molecular dipole configurations is presented in \Fig{dipoleorientation}a. This invariance can be explained by the dependence of cavity-induced intramolecular and intermolecular reaction pathways on $(\vec{\mu}_i(x_i)\cdot \vec{e})^2$, as detailed in Ref. \onlinecite{Ke_J.Chem.Phys._2024_p224704}. 

Furthermore, cavity-induced intermolecular reaction pathways differ from cavity-induced intramolecular reaction pathways, in that they do not require cavity thermalization with the cavity bath. Consequently, in a lossless cavity ($\lambda_{\mathrm{c}}=0$), cavity-induced rate modification arises solely from cavity-induced intermolecular reaction pathways, which require the overlap of vibrational transition energies in different molecules and the cavity frequency.

In a molecular ensemble, where each molecule has a distinct orientation, reaction rates vary between molecules.  \Fig{dipoleorientation}b illustrates the reaction rates as a function of $\lambda_{\mathrm{c}}$ for different molecules in a molecular ensemble with $N_{\mathrm{mol}}=4$. The first molecule is aligned along the light polarization, achieving the maximum light-matter coupling strength $\eta_{\mathrm{c}}$. The second and third molecules are oriented at angles of $\pi/6$ and $\pi/3$ to the light polarization, resulting in effective light-matter coupling strength per molecule of $\sqrt{3}\eta_c/2$ and $\eta_c/2$, respectively. These reduced coupling strengths lead to lower reaction rates and smaller turnover values of $\lambda_{\mathrm{c}}$. The fourth molecule, with its dipole moment perpendicular to the light polarization, is decoupled from the cavity and other molecules, showing no rate modification inside the cavity. Therefore, in a realistic molecular ensemble collectively coupled to the cavity mode, the disparity in reaction rates per molecule due to their distinct dipole orientations raises questions about how to define a group rate for such a system. This remains an open issue in rate theory and warrants further investigation.

\end{document}